\documentclass{aa}
\usepackage{psfig,latexsym,longtable,lscape}

\def\ros{{\sl ROSAT}}

\def\ein{{\sl Einstein}}

\def\HI{\hbox{H\,{\sc i}}}
\def\HII{\hbox{H\,{\sc ii}}}
\newcommand{\D}{$^\circ$}
\def\p0{\phantom{0}}

\newcommand\approxgt{\mbox{$^{>}\hspace{-0.24cm}_{\sim}$}}

\def\it{\sl}

\begin{document}
           
   \thesaurus{03         
              (09.19.2;  
               11.13.1;  
               13.25.2;  %
               13.25.5;  
               04.03.1)  %
             }
   \title{A ROSAT PSPC X-Ray Survey of the Small Magellanic Cloud}
 
   \author{P. Kahabka\inst{1,2},
           W. Pietsch\inst{3},
           M.D. Filipovi\'c\inst{3,4,5} and
           F. Haberl\inst{3}
          }

   \offprints{ptk@astro.uva.nl}
 
   \institute{$^1$~Astronomical Institute, University of Amsterdam, 
              Kruislaan 403, NL-1098 SJ Amsterdam, The Netherlands\\
              $^2$~Center for High Energy Astrophysics, University of
              Amsterdam, Kruislaan 403, NL-1098 SJ Amsterdam, The
              Netherlands\\
              $^3$~Max-Planck-Institut f\"ur extraterrestrische Physik,
              D--85740 Garching bei M\"unchen, Germany\\
              $^4$~University of Western Sydney, Nepean, P.O. Box 10,
              Kingswood, NSW 2747, Australia\\
              $^5$~Australia Telescope National Facility, CSIRO, P.O. Box
              76, Epping, NSW 2121, Australia
             }

   \date{Received 23 June 1998 / Accepted 7 December 1998}
 
   \maketitle\markboth{P. Kahabka et al.: A ROSAT PSPC X-ray Survey of the SMC}
                      {P. Kahabka et al.: A ROSAT PSPC X-ray Survey of the SMC}

   \begin{abstract}
We present the results of a systematic search for point-like and moderately
extended soft (0.1-2.4~keV) \mbox{X-ray} sources in a raster of nine 
pointings covering a field of $8.95\ {\rm deg}^2$ and performed with the \ros\ 
{\sl PSPC} between October~1991 and October~1993 in the direction of the 
Small Magellanic Cloud (SMC). We detect 248 objects which we include in 
the first version of our SMC catalogue of soft X-ray sources. We 
set up seven source classes defined by selections in the count rate, 
hardness ratio and source extent. We find five high luminosity super-soft 
sources (1E~0035.4-7230, 1E~0056.8-7146, RX~J0048.4-7332, RX~J0058.6-7146
and RX J0103-7254), one low-luminosity super-soft source RX~J0059.6-7138 
correlating with the planetary nebula L357, 51 candidate hard X-ray binaries
including eight bright hard \mbox{X-ray} binary candidates, 19 supernova remnants 
(SNRs), 19 candidate foreground stars and 53 candidate background active 
galactic nuclei (and quasars). We give a likely classification for $\sim$60\% of the 
catalogued sources. The total count rate of the detected point-like and 
moderately extended sources in our catalogue is $6.9\pm0.3$ s$^{-1}$, 
comparable to the background subtracted total rate from the integrated field 
of $\sim6.1\pm0.1$ s$^{-1}$. 
%
%
   \end{abstract}

%
%
\section{Introduction}
The SMC was the subject of several recent high resolution surveys in 
different wavelength bands. A CO survey of the SMC was produced by Rubio 
et al. (1993 a,b). Recent H$\alpha$ surveys were performed by Coarer et 
al. (1993) and Caplan et al. (1996) (see also Davies et al. 1976). At radio 
frequencies a neutral Hydrogen survey at 1.4~GHz (Staveley-Smith et al. 1997) 
and continuum surveys at 1.4 and 2.3~GHz (Filipovi\'c et al. in prep.) were 
made with the Compact Array of the Australia Telescope National Facility.
Also, Parkes radio surveys of the SMC were undertaken by Filipovi\'c et 
al. (1997). 

Early X-ray surveys were performed with the \ein\ satellite (Seward \& Mitchell 
1981; Inoue et al. 1983; Bruhweiler et al. 1987; Wang \& Wu 1992). Wang \& Wu 
(1992) found 20 sources from a sample of 70 sources to be intrinsic to the 
SMC. Filipovi\'c et al. (1998) compared the Parkes radio catalogue with 
the {\sl ROSAT PSPC} catalogue presented in this paper and found 27 sources 
common to these two surveys. 

Our aim in this paper is to present the first-pass catalogue of \ros\ X-ray 
sources in the SMC region with their basic properties and classification. 
This catalogue is given in Table~1. We classify about 60\% of the sources by 
making use of \mbox{X-ray} spectral parameters and correlations with optical, radio 
and other X-ray catalogues. Each source class is discussed separately in terms 
of statistical properties. 

\section{Observations and data analysis}
 
The observations used in this analysis were carried out with the {\it PSPC} 
detector on-board the {\it ROSAT} observatory during nine pointed observations
between 8~October~1991 and 14~October~1993. The satellite, its X-ray telescope
(XRT) and the focal plane detector ({\it PSPC}) used were discussed in detail 
by Tr\"umper (1983) and Pfeffermann et al. (1986). The pointings were 
performed in a raster covering the {\sl Optical Bar} and the {\it Eastern 
Wing} of the SMC quite homogeneously (cf. Fig.~1, Fig.~2 and Table~1 in 
Kahabka \& Pietsch 1996, herafter Paper~I). A search for unresolved and 
moderately extended sources has been conducted on the data obtained from 
these fields (Table~2).

\setcounter{table}{1}
\begin{table}[htbp]
  \caption[]{Fields used for source search in the SMC Survey.}
  \begin{flushleft}
  \begin{tabular}{llccr}
  \hline
  \noalign{\smallskip}
   Point. & Sequence     & RA          & Dec            & Expos.     \\
   (Pap.\,I) & Number  & (J2000.0)     & (J2000.0)      & (ksec)        \\
  \noalign{\smallskip}
  \hline
  \noalign{\smallskip}
  A1 & 600195p-0  & 0$^h$58$^m$12.0$^s$  & -72$^d$16$^m$48$^s$ & 16.6 \\
  A2 & 600195p-1  & 0$^h$58$^m$12.0$^s$  & -72$^d$16$^m$48$^s$ &  9.4 \\
  B1 & 600196p-0  & 0$^h$50$^m$12.0$^s$  & -73$^d$13$^m$48$^s$ &  1.3 \\
  B2 & 600196p-1  & 0$^h$50$^m$12.0$^s$  & -73$^d$13$^m$48$^s$ & 22.2 \\
  C  & 600197p    & 1$^h$13$^m$24.0$^s$  & -72$^d$49$^m$12$^s$ & 21.5 \\
  D  & 600452p    & 1$^h$05$^m$55.2$^s$  & -72$^d$33$^m$36$^s$ & 14.2 \\
  E  & 600453p    & 0$^h$54$^m$28.7$^s$  & -72$^d$45$^m$36$^s$ & 17.6 \\
  F1 & 600454p-0  & 0$^h$42$^m$55.2$^s$  & -73$^d$38$^m$24$^s$ &  9.7 \\
  F2 & 600454p-1  & 0$^h$42$^m$55.2$^s$  & -73$^d$38$^m$24$^s$ &  8.3 \\
  G1 & 600455p-0  & 1$^h$01$^m$16.7$^s$  & -71$^d$49$^m$12$^s$ &  3.6 \\
  G2 & 600455p-1  & 1$^h$01$^m$16.7$^s$  & -71$^d$49$^m$12$^s$ &  1.7 \\
  G3 & 600455p-2  & 1$^h$01$^m$16.7$^s$  & -71$^d$49$^m$12$^s$ &  4.6 \\
  G4 & 600455p-3  & 1$^h$01$^m$16.7$^s$  & -71$^d$49$^m$12$^s$ &  4.1 \\
  X0 & 400299p-0  & 0$^h$37$^m$19.2$^s$  & -72$^d$14$^m$24$^s$ &  5.1 \\
  X1 & 400299p-1  & 0$^h$37$^m$19.2$^s$  & -72$^d$14$^m$24$^s$ &  1.7 \\
  X2 & 400299p-2  & 0$^h$37$^m$19.2$^s$  & -72$^d$14$^m$24$^s$ &  2.3 \\
  Y1 & 400300p-0  & 0$^h$58$^m$33.5$^s$  & -71$^d$36$^m$00$^s$ &  5.2 \\
  Y2 & 400300p-1  & 0$^h$58$^m$33.5$^s$  & -71$^d$36$^m$00$^s$ &  7.2 \\
  Y3 & 400300p-2  & 0$^h$58$^m$33.5$^s$  & -71$^d$36$^m$00$^s$ &  7.1 \\
  \noalign{\smallskip}
  \hline
  \end{tabular}
  \end{flushleft}
\end{table}

A sophisticated detection procedure was applied to the SMC survey. Each
pointed observation has been analyzed with three detection methods (local, 
map and maximum likelihood, cf. Zimmermann et al. 1994). These detection 
procedures were applied to the data of single pointings given in Table~2.
Data with the same pointing direction have been merged to one data set. The
analysis was performed in the five  energy channel ranges Soft = (channel 
11-41, 0.1-0.4~keV), Hard = (channel 52-201, 0.5-2.1~keV), Hard1 = (channel 
52-90, 0.5-0.9~keV), Hard2 = (channel 91-201, 0.9-2.0~keV) and Broad 
(0.1-2.4~keV). The five source lists were merged to one final source list
taking only detections at an off-axis angle $\le45^{'}$ into account. This 
list comprises the source catalogue published in this paper. The maximum 
likelihood algorithm was used to determine the final source position, the 
counts in five energy bands and the source extent. A one-dimensional energy 
and position dependent Gaussian distribution (cf. Zimmermann et al. 1994) 
was applied in order to obtain the source extent. The source extent ($Ext$) 
is given as the Gaussian $\sigma_{\rm Gauss}$

\begin{equation}
  Ext = \sigma_{\rm Gauss} = FWHM_{\rm Gauss} / 2.35
\end{equation}

Hardness ratios HR1 and HR2 were calculated from the counts in the bands 
as \mbox {HR1=(H-S)/(H+S)} and \mbox {HR2=(H2-H1)/(H1+H2)}. The existence 
likelihood ratio and the extent likelihood ratio was calculated according 
to Cash (1979) and Cruddace et al. (1988). 

\section{The catalogue}

We selected for our final source catalogue only detections with an existence 
likelihood ratio \mbox{$\rm LH_{\rm exist}>10$}, which is equal to a probability of 
existence \mbox {$\rm P\sim(1-exp(-LH_{\rm exist}))\sim(1-4.5\times10^{-5})$}.
We give the value for the extent only in case the extent likelihood ratio is 
\mbox {$\rm LH_{\rm extent}\ge10$}.

A $90\%$ source error radius was calculated, adding quadratically
a $5\arcsec$ systematic error. 

\begin{equation}
  P_e = 2.1\times \sqrt{x_{\rm err}^2 + y_{\rm err}^2 + (5\arcsec)^2}
\end{equation}

The positional error derived for large off-axis angles \mbox {$\Delta 
\approxgt30'$} may be somewhat underestimated due to the asymmetry of the
point-spread-function. But the positional error should not be larger than
$\sim1^{'}$.

We catalogued all point-like and moderately extended sources found in 
this survey in Table~1. A small fraction 
($\sim$9\%) of catalogued sources may be false (artifacts due to the
PSPC detector structure). The source positions were not corrected with 
respect to the positions of known standard sources (e.g. foreground stars or 
background active galactic nuclei (AGNs)) because it may easily introduce improper shifts in position. 
Column~1 gives the source catalogue number and Col.~2 the {\sl ROSAT} 
source name. In Cols.~3 and 4 we list source positions, the right ascension 
(RA) and declination (Dec) for the epoch J2000 with the 90\% confidence 
positional uncertainty (Col.~5). Column~6 gives the total count rate (with the 
\mbox {$1\sigma$} errors). We list the soft (HR1) and the hard (HR2) hardness
ratio (with $1\sigma$ errors) in Cols.~7 and 8, while in Col.~9 we present 
the source extent in cases where the likelihood ratio for extent is greater 
than 10. In Col.~10 is listed the likelihood ratio of existence 
(LH$_{\rm exist}$) and in Col.~11 the off-axis angle ($\Delta$) for the 
detection with the highest LH$_{\rm exist}$ from different bands. Columns~12 
and 13 give the \ein~index (from the Wang \& Wu (1992) catalogue) and the 
distance to the \ein~source. Columns~14, 15 and 16 list the stellar type, the 
magnitude of the {\sl Simbad} identified star and the distance to the 
{\sl Simbad} star. In Col.~17 we list the source class according to our 
classification scheme (Sect.5, Table~3). In addition four sources are 
classified as SNRs although they have a smaller extent 
likelihood ratio as required from the classification. However, they coincide 
with SNRs detected with \ein . In Col.~18 we give a refined 
classification for candidate AGN and hard X-ray binary. In Col.~19 we give 
some notes on either radio or optical identifications. 

For the hard X-ray binary and the AGN class we have refined the classification
scheme. We have taken the local Hydrogen column into account and from 
simulations of power-law slope --2.0 and --2.6 AGN spectra we have predicted 
hardness ratios which we compared with the measured hardness ratios. In case 
the classification was not unique we introduced the class AB in Table~1,
column~18.

In Fig.~1 we show the spatial distribution of all 248 detections in the 
direction of the SMC overlaid on a merged exposure image. Only the central 
$45^{'}$ of the field of view are considered for source detection.

\begin{figure*}
  \resizebox{13.5cm}{!}{\includegraphics{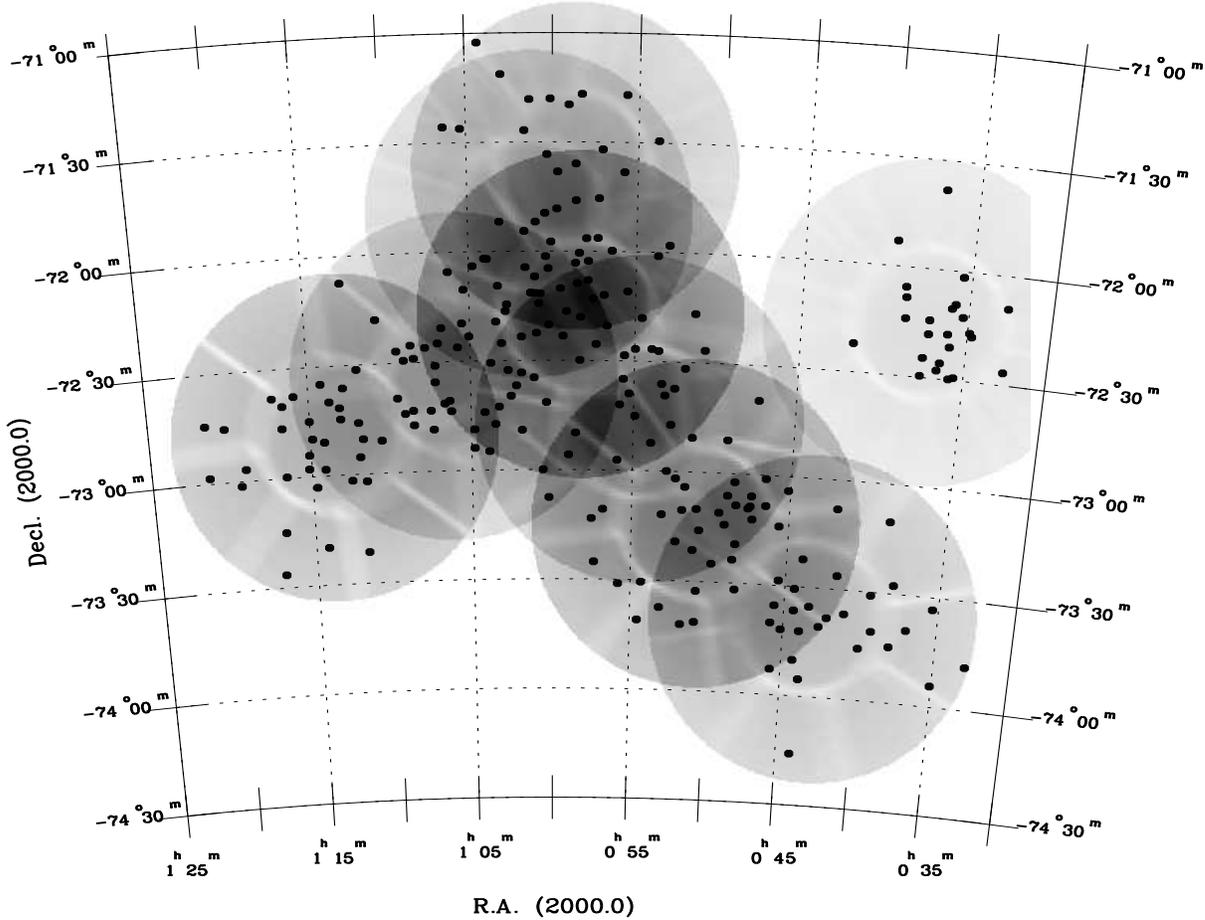}}
    \caption{Distribution of all X-ray detections in the field of the SMC
             drawn over integrated exposure map.}
    \label{label2}
\end{figure*}

\section{The survey}

The X-ray survey of the SMC covers a field of $8.95\ {\rm deg}^2$ considering
the used $45^{'}$ field of view of each pointed observation. From the merged
(0.1-2.4)~keV image corrected for exposure we derive a total count rate 
of 53.2 s$^{-1}$. The mean exposure of this image is 7\,500~s and the
maximum 22\,000~s. We created a merged background image of the
same field and subtracted the background rate image from the count rate 
image. We find an excess rate of $6.1\pm0.1$ s$^{-1}$. In comparison the total 
source count rate we derive from our catalog of 248 point-like and moderately 
extended sources is $6.9\pm0.3$ s$^{-1}$. Both rates are within the 
2$\sigma$ errors identical and are in agreement with the finding from the 
\ros\ All-Sky Survey (RASS) data that the residual count rate of an extended emission component 
(a hot gas) is small. Kahabka \& Pietsch (1993) find an excess count rate 
of $\sim 0.7$ s$^{-1}$ for the $8\times8\ {\rm deg}^2$ RASS field after 
subtracting a smooth background component and the source contribution. 
This result may be compared with the finding of Irwin \& Sarazin (1998). They
found that the integrated X-ray emission and colors of X-ray faint galaxies 
(group~1 galaxies) to which also M31 and the SMC are belonging can be 
explained as the integrated emission from (low-mass) X-ray binaries.

\section{Source classes}

The X-ray spectral characteristics of the catalogued sources are given as two 
hardness ratios (the soft hardness ratio - HR1 and the hard hardness ratio - 
HR2), the source extent (assuming a Gaussian model) and the count rates in 
a spectrally soft and hard band. This information is used to characterize a 
source and to allow a source classification. We previously applied some 
selections to this catalogue in order to derive the sample of super-soft \mbox{X-ray}
sources of the SMC (Kahabka et al. 1994). In Kahabka \& Pietsch (1996) we 
applied another set of selections in order to derive the sample of spectrally 
hard X-ray binary candidates with luminosities in excess of \mbox {$\rm
3\times10^{35}\ {\rm erg}\ {\rm s}^{-1}$}. These subsets comprise only a 
small number of sources (four supersoft and 7-13 hard). The majority of 
sources (209) have not been classified previously. 

In a recent paper Filipovi\'c et al. (1998) presented results of a comparison 
between the {\sl ROSAT PSPC} catalogue presented here and the Parkes
catalogue of sources towards the SMC at radio frequencies (1.42, 2.45, 4.75, 
4.85 and 8.55~GHz). They found 27 sources in common to both surveys (cf. 
Table~3 in their paper). These include 14 SNRs (with two SNR candidates), 
eight background sources and three \HII\ regions.

We now define other classes, set up the selection criteria, identify the 
members, derive the relevant distributions of the source properties and 
discuss the implications in terms of population studies. The positional 
distribution of the classified sources using criteria established here 
(Table~3) is shown in Fig.~2.

\setcounter{table}{2}
\begin{table*}[htbp]
  \caption[]{Definition of the source classes and the selection criteria}
  \begin{flushleft}
  \begin{tabular}{llcccc}
  \hline
  \noalign{\smallskip}
  \multicolumn{2}{c}{Source Class} & \multicolumn{4}{c} {Selection Criteria}                                      \\
  \multicolumn{2}{c}{}                  & Count Rate & HR1                    & HR2               & Extent \\ 
  \multicolumn{2}{c}{}                  & (s$^{-1}$) &                        &                   & Likelihood\\
  \noalign{\smallskip}
  \hline
  \noalign{\smallskip}
Sl & High luminosity super-soft sources & $>$0.015   & HR1+$\delta$HR1$<-$0.8 &                        &      \\
Sw & Low luminosity super-soft sources  & $<$0.015   & HR1+$\delta$HR1$<-$0.8 &                        &      \\
Bl & Stronger hard X-ray binaries       & $>$0.015   & HR1--$\delta$HR1$>$+0.5 & HR2--$\delta$HR2$>$+0.3 & $<$50\\
Bw & Weaker hard X-ray binaries         & $<$0.015   & HR1--$\delta$HR1$>$+0.5 & HR2--$\delta$HR2$>$+0.3 & $<$50\\
R  & SNRs and extended structure        &            & HR1--$\delta$HR1$>-$0.8 &                        & $>$50\\
F  & Foreground stars                   &            & HR1+$\delta$HR1$<$+0.5 & HR2--$\delta$HR2$>-$0.8 & $<$50\\
A  & Background objects (AGNs)          &            & HR1--$\delta$HR1$>$+0.5 & HR2+$\delta$HR2$<$+0.3 & $<$50\\
D  & Possible artifacts                 &            &                        &                        &      \\
  \noalign{\smallskip}
  \hline
  \noalign{\smallskip}
  \end{tabular}
  \end{flushleft}
Remark: additional classes from refined classification: AB = AGN or hard X-ray binary source and
H = \HII\ region. 
\end{table*}

\begin{figure*}
  \resizebox{13.5cm}{!}{\includegraphics{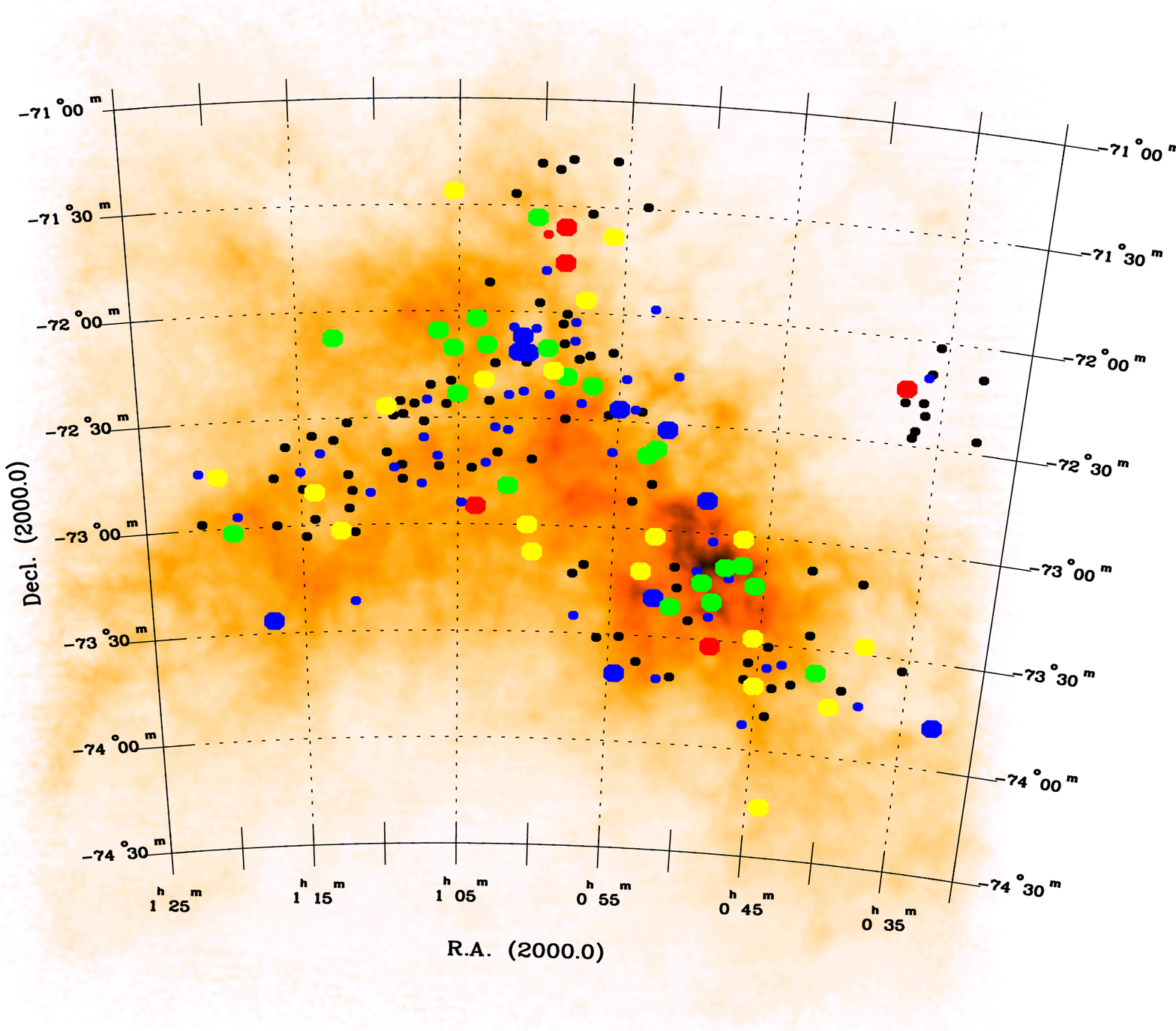}}
    \caption{Distribution of classified X-ray detections (this work) in the 
             field of the SMC plotted over high resolution \HI\ image of the
             SMC (from Stanimirovic et al. 1998). Selected classes are: 
             super-soft sources (red), hard X-ray binaries (blue), AGNs 
             (black), SNRs (green) and foreground stars (yellow). Big/small 
             circles are drawn for the high/low luminosity subclass 
             respectively. For the AGNs and hard X-ray binaries the refined
             classification is used (cf. Tab.1).}
    \label{label2}
\end{figure*}

\subsection{The high luminosity super-soft sources}

This class was extensively discussed in Kahabka et al. (1994) and in 
Paper~I. Applying the strict selection criteria outlined in these papers 
(\mbox {HR1$+\delta$HR1$<-0.8$} and count rate \mbox {$\rm >0.015~s^{-1}$}) 
we obtain candidates classified as class=Sl in Table~1. Sources No.~20, 62, 
133, 135 and 181 were discussed in Paper~I. Sources No.~6, 112 and 173 
coincide with struts of the {\sl PSPC} detector system and are classified as 
possible detector artifacts.

\subsection{The low luminosity super-soft sources}

Low luminosity super-soft sources are not yet established as a new class of
objects, but they are expected to exist e.g. as hot central stars of
planetary nebulae (PNe) which, if they are on the cooling track, can attain
luminosities in the range $\rm 10^{34} - 10^{36} {\rm erg}\ {\rm s}^{-1}$. 
This would translate into {\sl PSPC} count rates for sources in the SMC 
of $\rm \sim7\times10^{-5} - 7\times10^{-3}\ s^{-1}$ i.e. below our threshold 
of \mbox {$\rm 0.015\ s^{-1}$} for high luminosity super-soft sources. 

There are three objects which are fulfilling these criteria and they are 
classified as class=Sw in Table~1. The source No.~38 is found at a large 
off-axis angle of 43$'$ and is extended (Gaussian $\sigma=214''$). This may 
indicate either diffuse structure or an artifact. The source No.~141 
coincides with a {\sl Simbad} star of 17.5~mag and with the planetary nebula 
L357 in the catalogue of Meyssonnier \& Azzopardi (1993). It is therefore a 
real candidate. Source No.~243 is found at an off-axis angle of $20'$ and 
coincides with the inner {\sl PSPC} support structure and may be an artifact. 
This leaves us with one firm low-luminosity super-soft source No.~141 
(RX~J0059.6-7138).

\subsection{The ``stronger'' hard X-ray binary candidates} 

The sample of X-ray binary candidates with luminosities in excess of
\mbox {$\rm 3\times10^{35}\ {\rm erg}\ {\rm s}^{-1}$} has been given in 
Paper~I. The selection criteria were \mbox{HR1--$\delta$HR1$>0.5$}, 
\mbox{HR2--$\delta$ HR2$>0.3$} with extent likelihood \mbox {$\rm LH_{\rm 
extent}<50$} and count rate \mbox {$\rm >0.015~s^{-1}$}. We classify these 
sources as class=Bl in Table~1. 

The sources No.~3, 69, 83, 84, 100, 157, 158, 160 and 242 were selected as 
candidate X-ray binaries. In Paper~I sources with No.~26, 105 and 153 were 
rejected mainly due to the absence of time variability. For these
sources radio emission was found in the radio survey of Filipovi\'c et 
al. (1998) and they are candidates for AGNs.

The source No.~103 (RX~J0054.9-7226) is identified with the XTE~J0055-724 = 
1SAX~J0054.9-7226 source (Marshall et al. 1998, Israel 1998). The detection 
of pulsations with a period of 59~s with {\sl BeppoSAX} confirm the X-ray 
binary nature of this source. 

Stevens et al. (1998) identified early type emission-line stars through colour 
indices and H$\alpha$ emission for the sources with catalogue indices 3, 69, 
103, and 158.   

The source No.~153 (RX~J0100.7-7211) was considered to be consistent 
with a background AGN shining through the SMC bulge (Paper~I). Sources No.~157 
and 160 were found to coincide with detector struts and were rejected 
accordingly (Paper~I). 

\begin{table}[!]
  \caption[]{New X-ray pulsars discovered in recent {\sl ASCA}, {\sl 
BeppoSAX} and  {\sl Rossi-XTE} observations of the SMC.}
  \begin{flushleft}
  \begin{tabular}{rlccl}
  \hline
  \noalign{\smallskip}
Cat.  &  Source name              & Pulsation          & Period    & Ref. \\
No.   &                           & Period [s]         &       [d] &      \\
  \noalign{\smallskip}
  \hline  
  \noalign{\smallskip}
 98 & XTE J0053-724               & 46.64(4)           & 139?      & 1-3  \\
    & \hspace{-3.1mm}= 1WGA J0053.8-7226         &                    &           &      \\
  \noalign{\smallskip}
 79 & AX J0051-722                & 91.12(5)           & $\sim$120 & 1,4,5\\
  \noalign{\smallskip}
    & XTE J0054-720               &169.3-168.4         & -         & 1    \\
  \noalign{\smallskip}
103 & XTE J0055-724               & 59.072(1)          & -         & 5-7  \\
    & \hspace{-3.1mm}= 1SAX J0054.9-7226         &                    &           &      \\
  \noalign{\smallskip}
 69 & AX J0049-729                & 74.8(4)            & -         & 8,9  \\
    & \hspace{-3.1mm}= RX J0049.1-7250           &                    &           &      \\
  \noalign{\smallskip}
 72 & AX J0051-73.3               & 323.2(4)           &$^{*)}$0.708 & 10-13\\
    & \hspace{-3.1mm}= RX J0050.7-7316           &                    &           &      \\
  \noalign{\smallskip}
    & AX J0058-72.0               &280.4(3)            & -         & 14   \\
  \noalign{\smallskip}
    & 1SAX J0103.2-7209           &345.2(1)            & -         & 5    \\
    & \hspace{-3.1mm}= AX J0103-722              &                    & -         & 15   \\
  \noalign{\smallskip}
    & AX J0105-722                &3.34300(3)          & -         & 21   \\
  \noalign{\smallskip}
    & AX J0049-732                &9.1321(4)           & -         & 22   \\
  \noalign{\smallskip}
    & XTE J0111.2-7317            &31.0294(7)          & -         & 23-25\\
  \noalign{\smallskip}
  \hline
  \noalign{\smallskip}
  \multicolumn{5}{c}{Previously known SMC X-ray pulsars}                  \\
  \noalign{\smallskip}
  \hline
  \noalign{\smallskip}
  242 & SMC X-1                   & 0.7077             & 3.892     & 16   \\
  \noalign{\smallskip}
      & RX J0059.2-7138           & 2.76               & -         & 17   \\
  \noalign{\smallskip}
      & RX J0117.6-7330           & -                  & -         & 18   \\ 
  \noalign{\smallskip}
   83 & RX J0051.8-7231           & 8.9                & -         & 19   \\
      & \hspace{-3.1mm}= 2E 0050.1-7247          &                    &           &      \\
   \noalign{\smallskip}
  \hline
  \noalign{\smallskip}
  \end{tabular}
  \end{flushleft}
 Ref.  Corbet et al. 1998 (1), Lochner 1998b (2), Buckley et al. 1998a (3),
   Lochner 1998a (4), Israel et al. 1998 (5), Marshall et al. 1998 (6), 
   Israel 1998 (7), Yokogawa \& Koyoma 1998a (8), Kahabka \& Pietsch 1998 (9), 
   Yokogawa \& Koyoma 1998b (10), Kahabka 1998 (11), Cook 1998 (12), Schmidtke
   \& Cowley 1998 (13), Yokogawa \& Koyama 1998b (14), Yokogawa \& Koyoma
   1998c (15), Wojdowski et al. 1998 (16), Hughes 1994 (17), Clark et al. 1997
   (18), Israel et al. 1997 (19), Buckley et al. 1998b (20), Yokogawa \& 
   Koyama 1998d (21), Imanishi et al. 1998 (22), Chakrabarty et al. (1998a) 
   (23),  Wilson \& Finger (1998) (24), Chakrabarty et al. (1998b) (25). \\
   $^{*)}$ MACHO has seen a 0.708 day period. This is consistent with a rotation 
   period of a Be star.
\end{table}

\subsection{The ``weaker'' hard X-ray binary candidates}     

Weak hard X-ray binaries are an interesting class of \mbox {X-ray} objects
as they have been predicted to exist and their number is expected to be
large especially in galaxies of low metallicity like the SMC. In previous
work (Bruhweiler et al. 1987; Wang \& Wu 1992) candidates for such sources 
have been found and either classified as low luminosity Be systems or 
as background objects. 

Here, we are searching for candidates of this class by applying the same 
selection criteria as for the strong (or higher luminosity) hard X-ray 
binaries as outlined in Paper~I: \mbox {HR1$-\delta$HR1 $>0.5$}, 
\mbox {HR2$-\delta$HR2$>0.3$} and extent likelihood \mbox {$\rm LH_{\rm 
extent}<$50}. The only difference is to select objects with count rates of 
\mbox {$\rm <0.015\ s^{-1}$}, i.e. with luminosities below \mbox {$\rm \sim 
3\times10^{35}\ {\rm erg}\ {\rm s}^{-1}$} assuming a standard spectral model 
for the source flux (Paper~I). There are 60 such objects and we tentatively 
classify these sources as class=Bw. This class is a substantial fraction 
(25\%) of the total catalogue entries and turns out to be the class with most 
members. 

We find 15 of these objects which coincide with \ein\ detections. This may 
reflect that we are considerably deeper in sensitivity than the \ein\ 
survey. Some 17 objects were found to coincide with a {\sl Simbad} source. 
However, such a correlation could be misleading as most of the distances to 
the {\sl Simbad} sources are too large ($>60''$) to be considered reliable. 
Only three sources have a distance to a {\sl Simbad} source of $<50''$ and 
may be considered to be identified. This is a very small fraction of all 
catalogued sources here.

Four sources are found close to the inner ring of the {\sl PSPC} window 
support system and may be artifacts. A more thorough investigation of 
these sources appears to be required as these sources can still be real. 
Another 46 sources have been found inside the inner support ring of the 
{\sl PSPC} detector and are considered as firm candidates. This comprises 
22\% of all reliable entries in the catalogue. Sources found outside the 
detector ring suffer due to less accurate positions.

``Screening'' of the catalog using hardness ratios derived from simulated 
power-law slope --0.8 spectra for hard X-ray binaries, and power-law slope --2.0, 
and slope --2.6 spectra for ``radio loud'' and ``radio quiet'' AGNs respectively 
gives 43 firm weak hard X-ray binary candidates. Six previous hard X-ray binary 
candidates are consistent with AGNs and 11 candidates are consistent with 
either class (cf. Table~1). 

Recent observations towards the SMC with {\sl ASCA}, {\sl BeppoSAX} and 
{\sl Rossi-XTE} established 10 new X-ray pulsars in this galaxy (Tab.4). 
Most (if not all) of them appear to be connected with a Be-type donor star. 
Pulsation periods in the range 3-345 s have been determined. This 
range in pulsation periods is covered by the range of pulsation periods 
found in the galactic Be-star X-ray binaries of $\sim$4--1500 s (cf. 
van den Heuvel \& Rappaport 1987). An orbital period has been estimated only 
for the two systems AX~J0051-722 and XTE~J0053-724 with 110-120 and 139 days. 
The orbital period deduced for AX~J0051-722 is in agreement with the relation 
between pulsation period and orbital period found by Corbet (1986), while the
orbital period estimated for XTE~J0053-724 is twice the predicted period.
Five of these new X-ray pulsars may have a counterpart in our SMC X-ray 
catalogue.
Source~98 (RX~J0053.9-7226) has been discovered with {\sl Rossi-XTE
= RXTE} (cf. Levine et al. 1996) in an outburst and 46.6~s pulsations have 
been found (Corbet et al. 1998). This confirms the correct classification of 
this source. In addition, source~79 (RX~J0051.3-7216) may be identical with 
AX~J0051-722 (Corbet et al. 1998) also confirming the correct classification. 
Source~89 coincides with the transient source \mbox{RX~J0052.9-7158} of 
Cowley et al. (1997) and AX~J0051-73.3 (Yokogawa et al. 1998b) coincides with 
RX~J0050.7-7316 (cf. Cook 1998; Cowley et al. 1997; Kahabka 1998). This
source with catalog index 72 apparently both fits to the AGN and the hard
X-ray binary class but it is a strong candidate for a hard X-ray binary.

The derived number of 51 hard X-ray binary candidates may be compared with 
the number of X-ray binaries predicted from the population synthesis calculations 
of Dalton and Sarazin (1995) for the SMC. These calculations predict 46 \mbox{X-ray} 
binaries with luminosities in excess of $10^{34}\ {\rm erg}\ {\rm s}^{-1}$, 
the lower sensitivity limit of our SMC X-ray survey.  

\begin{table}[!]
  \caption[]{Upper panel: Previously detected SNRs identified in the \ros\ sample, 
             classified as SNR candidates and extended structure. The \ein\ 
             No. refers to Wang \& Wu (1992). Source radio name is given following 
             Filipovi\'c et al. (1998).}
  \begin{flushleft}
  \begin{tabular}{cccc}
  \hline
  \noalign{\smallskip}
Cat.  & \ein & Name of SNR    & Radio           \\
No.   & No.  &                & Source Name     \\
  \noalign{\smallskip}
  \hline  
  \noalign{\smallskip}
 34   &  -   & -              & SMC B0039-7353 \\
 55   & 15   & SNR~0044-7325  & -              \\
 61   & 16   & SNR~0045-734   & SMC B0045-7324 \\
 68   & 22   & SNR~0047-735   & SMC B0047-7332 \\
 77   & 24   & SNR~0049-736   & SMC B0049-7338 \\
140   & 44   & SNR~0057-724   & SMC B0057-7226 \\
189   & 52   & SNR~0103-726   & SMC B0103-7239 \\
191   & 53   & DEM S128       & SMC B0104-7226 \\
195   & 54   & SNR~0104-723   & - \\
  \noalign{\smallskip}
  \hline
  \noalign{\smallskip}
\multicolumn{4}{c}{SNRs missed by our selection} \\
  \noalign{\smallskip}
  \hline  
  \noalign{\smallskip}
 63      & 21       & SNR~0046-735  & SMC~B0046-7333 \\
 86      & 30       & SNR~0050-728  & -              \\
 90      & -        &               & SMC~B0051-7254 \\
128      & 42       & SNR~0056-725  & -              \\
148      & -        & SNR~0058-718  & SMC~B0058-7149 \\
177      & 50       & SNR~0101-724  & SMC B0101-7226 \\
182/183  & 51 & SNR~0102-723  & SMC B0102-7218 \\
  \noalign{\smallskip}
  \hline  
  \end{tabular}
  \end{flushleft}
\end{table} 

\subsection{Supernova remnants and other extended structures}

SNRs have been identified in the SMC by work done in the radio, optical and 
X-ray regime (Mills et al. 1982, 1984). In the work of Ye \& Turtle (1993), 
some 15 SNRs and SNR candidates are detected in a 843~MHz survey. 

We applied the selection criteria likelihood of extent \mbox 
{$\rm LH_{\rm extent} >50$} and \mbox {HR1$-\delta$HR1$>-0.8$} in order to 
derive a candidate sample of SNRs and other extended structures. We find 19 
objects fulfilling these criteria (including four SNRs detected by \ein, 
although they have a LH$_{\rm ext}$ below our classification threshold: 
sources 86, 128, 177 and 182). These sources are classified as class=R 
in Table~1. The sources with catalogue number 95, 132, and 136 may be detector 
artifacts. They have the class=R/D. 
For 13 sources an \ein\ identification has been found. We give the \ein\ 
number and the identification for these sources in Table~5. 12 of these 
sources actually are known SNRs. The bright young oxygen-rich SNR~0102-723
(Amy \& Ball 1993) has two entries (182 and 183) in our catalog due to the merging of
two pointed observations at different off-axis-angles. Entry~183 is the
more accurate one. An additional classified source (34) correlates with 
a SNR proposed by Filipovi\'c et al. (1998). We find new candidate SNRs in our X-ray
survey: RX~J0101.8-7249 (source 165) and RX~J0112.7-7207 (223) were not 
reported before, while RX~J019.4-7301 (245) was already detected with 
\ein\ (BKGS~30, Bruhweiler et al. 1987), however not classified as SNR. 

We miss a few well known SNRs in the SMC with this selection. They have an 
extent likelihood ratio $>$10 and fulfill the criteria of being extended, 
suggesting we have chosen too strict an extent criterion in order to be on 
the secure side. It also suggests that there are still unrecognized SNRs in 
our sample (see discussion in Filipovi\'c et al. 1998). We list these 
SNRs in the lower part of Table~5.

\subsection{Foreground stars}

It is not trivial to select this sample just from the X-ray characteristics. 
Stars are coronal emitters with temperatures in the range of a few 
\mbox {$\rm10^6$} to \mbox {$\rm10^7$~K}. The HR1 would then fall into the 
regime \mbox {HR1$+\delta$HR1$<0.5$}, \mbox {HR2$-\delta$HR2$>-0.8$} and the 
likelihood of extent \mbox {$\rm LH_{\rm extent}<50$}. Actually, all galactic 
foreground stars detected in a \mbox {$8^{\circ}\times8^{\circ}$} field 
centered on the SMC and observed during the RASS 
have values of \mbox {HR1$ <$0.1} (Kahabka \& Pietsch 1993). This means that 
we may be too conservative in selecting stars in our sample with the criteria 
mentioned above. 

We find 19 candidates and we classify these sources as class=F in Table~1. 
For seven objects, a {\sl Simbad} match exists. Six of these identifications 
appear to be reliable as the distance to the {\sl Simbad} source is $<60''$. 
Five of them correlate with stars of spectral type O, A or F and they are 
given in Table~6. Source~138 coincides with an O star. Assuming a conversion 
factor of $0.76\times10^{37}\ {\rm erg}\ {\rm cts}^{-1}$ we derive for the
O-star in the SMC an X-ray luminosity of $10^{34}\ {\rm erg}\ {\rm s}^{-1}$ 
from the measured count rate. The nature of the other 14 sources remains 
unclear.

\begin{table}[!]
  \caption[]{Optically identified X-ray selected candidates identified
             with {\sl Simbad} stars}
  \begin{flushleft}
  \begin{tabular}{cccc}
  \hline
  \noalign{\smallskip}
Cat.  & Spectral & mag   & Distance \\
No.   & Type     &       & ($''$)   \\
  \noalign{\smallskip}
  \hline  
  \noalign{\smallskip}
   33 & F3V      &  8.9  & \p03     \\
   44 & F6IV     &  8.0  & 20       \\
  109 & FV/FVI   & 11.5  & 12       \\
  138 & O7III    & 14.1  & 52       \\
  246 & A9/F0    &  7.8  & 26       \\
  \noalign{\smallskip}
  \hline  
  \noalign{\smallskip}
  \multicolumn {4}{c}{Star identified in class A} \\
  \noalign{\smallskip}
  \hline  
  \noalign{\smallskip}
  179 & -        & 16.6  & 47      \\
  \noalign{\smallskip}
  \hline
  \end{tabular}
  \end{flushleft}
\end{table}

\subsection{Background objects and AGNs}

Background objects are selected as \mbox {HR1$-\delta$HR1$>0.5$}, 
\mbox {HR2$+\delta$HR2$<0.3$} and extent likelihood \mbox 
{$\rm LH_{\rm extent}<50$}. 
The same criterion has already been applied in Paper~I in order to identify 
a possible sample of AGNs in the high luminosity X-ray binary candidate sample. 
We find 20 sources to fulfill these criteria. Source~148 correlates with a 
SNR and has to be removed from this class. Three sources correlate within a 
radius of $<60''$ with a {\sl Simbad} star (either foreground or SMC).
We refine and extend the classification of candidate AGNs by comparing the 
measured hardness ratios with the predictions from simulated power-law spectra 
of slope --2.0 and --2.6. We find 53 candidates for class=A and 62 candidates
if we also consider class=AB in Table~1. Class=AB means a hard X-ray binary 
nature is also possible due to the hardness ratio criteria. 

Tinney et al. (1997) present 10 quasars behind the SMC. Seven of them are 
covered by the fields X1 and C (cf. Tab.\,2) and are listed in Tab.~7. None of 
these quasars was detected in the radio survey of Filipovi\'c et al. 
(1998). QJ0102-7546, one of the three quasars not covered by our survey 
was detected in the RASS field (Kahabka \& Pietsch 1993). 

\begin{table}[!]
  \caption[]{``Tinney'' quasars found in pointings X1 and C of the SMC X-ray 
             survey.}
  \begin{flushleft}
  \begin{tabular}{ccc}
  \hline
  \noalign{\smallskip}
Cat. No. & {\sl ROSAT} name   & Object        \\
  \noalign{\smallskip}
  \hline  
  \noalign{\smallskip}
  9 & RX~J0035.5-7201 & QJ0035-7201   \\
 16 & RX~J0036.5-7225 & QJ0036-7225   \\
 17 & RX~J0036.6-7227 & QJ0036-7227   \\
 20 & RX~J0037.3-7214 & QJ0037-7218   \\
218 & RX~J0111.7-7250 & QJ0111-7246   \\
224 & RX~J0112.8-7236 & QJ0112-7236   \\
238 & RX~J0116.5-7259 & QJ0116-7259   \\
  \noalign{\smallskip}
  \hline  
  \end{tabular}
  \end{flushleft}
\end{table} 

We independently calculate the number of background sources in the analyzed
field by taking the distribution of the neutral hydrogen into account. By 
making use of the standard log(N)--log(S) of the soft extragalactic X-ray 
background (Hasinger et al. 1993) and by taking into account the absorption 
due to the SMC (from the \HI\ image of Stanimirovic et al. 1998) the expected 
number of background sources with absorbed fluxes in excess of $10^{-13}\ 
erg\ {\rm cm}^{-2}\ {\rm s}^{-1}$ has been determined. We derive a number of 
10 background sources for our covered field. As we expect to be complete for 
this flux limit we consider 10 background sources as the lower limit. The 
lower end of the flux distribution extends to $10^{-14}\ {\rm erg}\ 
{\rm cm}^{-2}\ {\rm s}^{-1}$. In case of completeness we expect to detect 519 
background sources. The fact that we classify 53-62 sources as background 
sources is consistent with these numbers.

\section{Conclusions}
We performed an X-ray survey of a 8.95~{\rm deg}$^2$ field in the direction of
the Small Magellanic Cloud. We detect 248 point-like and moderately extended 
sources. Using criteria established here, six sources were classified as supersoft sources, 51 as hard X-ray binary
candidates, 19 as supernova remnants, 19 as candidate foreground stars and 53
as candidate background AGNs. These are 60\% of all catalog entries. The 
number of hard X-ray binaries agrees with the numbers predicted from 
population synthesis calculations for luminosities in excess of $10^{34}\ 
{\rm erg}\ {\rm s}^{-1}$. Assuming the standard log(N)-log(S) of the soft 
extragalactic X-ray background we estimate that in our field are 10 background
AGNs with fluxes in excess of $10^{-13}\ {\rm erg}\ {\rm cm}^{-2}\ 
{\rm s}^{-1}$ and 519 background AGNs with fluxes in excess of $10^{-14}\ 
{\rm erg}\ {\rm cm}^{-2}\ {\rm s}^{-1}$. We propose three new SNR 
candidates.

\acknowledgements
This research was supported in part by the Netherlands Organization for
Scientific Research (NWO) through Spinoza Grant 08-0 to E.P.J. van den Heuvel.
P.K. thanks E.P.J. van den Heuvel for stimulating discussions and X. Li 
for reading the manuscript. S. Stanimirovic is thanked for providing the
\HI\ image of the SMC. Part of the work has been performed during the stay of 
P.K. at the Max-Planck-Institut f\"ur extraterrestrische Physik in Garching.
The \ros\ project is supported by the Max-Planck-Gesellschaft and the 
Bundesministerium f\"ur Forschung und Technologie (BMFT). This research made 
use of the Simbad data base operated at CDS, Strasbourg, France. We thank the 
referee for useful comments and suggestions to improve this work.

\clearpage
\onecolumn
\pagestyle{empty}
\landscape
\noindent {{\bf Table~1.} \ros\ {\sl PSPC} Point Source Catalogue of the SMC
\scriptsize
  \begin{longtable}[l]{rcccccccrrrcrrcclll}
            \noalign{\smallskip}
            \hline 
            \noalign{\smallskip}
(1) &(2)          &(3)                 &(4)                                 &(5)          &(6)                                       &(7)             &(8)                &(9)         &(10)   &(11)         &(12)  &(13)                   &(14)  &(15)   &(16)         &(17)   &(18) &(19) \\
    & Source      & RA                 & Dec                                &             & \multicolumn{1}{c}{Count Rate}              & \multicolumn{2}{c}{Hardness Ratio} &             &       &             & \multicolumn{2}{c}{Einstein} & \multicolumn{3}{c}{Simbad} &       &     \\
 No.& Name        & (J2000)            & (J2000)                            & $\rm P_{\rm e}$ & Total                                  & HR1            & HR2               & Ext         & LH    & $\Delta$    & ID   &  Dist                 & Type & Mag   & Dist        &Class &Class & Note\\
    & RX~J        & \p0h~\p0m~\p0s\p0~ & \p0\D\p0\p0$\arcmin$~\p0$\arcsec$  & ($\arcsec$) & \multicolumn{1}{c}{$\rm (10^{-3}\ s^{-1})$} &                &                   & ($\arcsec$) &       & ($\arcmin$) &      & ($\arcsec$)           &      &       &($\arcsec$)  &   &(refined)&     \\
            \noalign{\smallskip}
            \hline
            \noalign{\smallskip}
  1 & 0032.6-7226 & 00 32 37.6         & -72 26 32                          & 34          & 3.86$\pm$1.12                               & 0.86$\pm$0.51  & 0.14$\pm$0.32    &   & 12&25& & & & && &A  &     \\
  2 & 0032.7-7208 & 00 32 42.3         & -72 08 59                          & 17          & 15.6$\pm$2.1                                & 0.79$\pm$0.14  & 0.11$\pm$0.14    &   &107&22& & & & &&A &  &     \\
  3 & 0032.9-7348 & 00 32 55.1         & -73 48 11                          & 16          & 118$\pm$5                                   & 1.0            & 0.55$\pm$0.04    & 45&909&43& & & &14.1&68&Bl& & \\
  4 & 0034.7-7217 & 00 34 43.3         & -72 17 42                          & 15          & 2.76$\pm$0.72                               & 1.0            & 0.18$\pm$0.26    &   & 31&12& & & & && &   &     \\
  5 & 0034.8-7216 & 00 34 51.9         & -72 16 44                          & 17          & 2.17$\pm$0.92                               & 0.67$\pm$0.51  & 0.56$\pm$0.31    &   & 15&11& & & & && &   &     \\
            \noalign{\smallskip}
  6 & 0035.0-7354 & 00 35 04.8         & -73 54 14                          & 35          & 21.0$\pm$4.80                               & -1.0           & 0.00$\pm$0.00    &170&104&36& & & & &&Sl/D& &     \\
  7 & 0035.3-7212 & 00 35 20.8         & -72 12 34                          & 13          & 4.87$\pm$0.90                               & 0.83$\pm$0.20  & 0.29$\pm$0.19    &   & 68& 9& & & & && &   &     \\
  8 & 0035.3-7333 & 00 35 23.3         & -73 33 19                          & 51          & 3.43$\pm$1.03                               & 1.0            & -0.08$\pm$0.34   &   & 10&32& & & & && A&  &     \\
  9 & 0035.5-7201 & 00 35 31.2         & -72 01 34                          & 17          & 6.32$\pm$1.42                               & 0.86$\pm$0.25  & 0.21$\pm$0.21    &   & 47&15& & & & && &A & quasar\\
 10 & 0035.6-7229 & 00 35 36.4         & -72 29 20                          & 32          & 8.25$\pm$1.86                               & 0.98$\pm$0.29  & 0.46$\pm$0.18    &   & 29&17& & & & && &   &     \\
            \noalign{\smallskip}
 11 & 0035.8-7209 & 00 35 51.0         & -72 09 13                          & 16          & 3.42$\pm$1.13                               & 0.22$\pm$0.36  & 0.47$\pm$0.30    &   & 19& 8& & & & && &A &     \\
 12 & 0035.8-7229 & 00 35 53.7         & -72 29 47                          & 25          & 4.38$\pm$1.11                               & 1.0            & -0.26$\pm$0.25   &   & 22&17& & & & &&A &   &     \\
 13 & 0036.0-7221 & 00 36 00.1         & -72 21 05                          & 12          & 6.89$\pm$1.22                               & 0.55$\pm$0.18  & 0.23$\pm$0.18    &   & 83& 9& & & & && &A &     \\
 14 & 0036.0-7210 & 00 36 02.6         & -72 10 24                          & 18          & 0.95$\pm$0.44                               & 1.0            & 1.0              &   & 11& 7& & & & &&Bw& &     \\
 15 & 0036.1-7217 & 00 36 10.8         & -72 17 34                          & 14          & 3.14$\pm$1.13                               & 0.31$\pm$0.31  & 0.28$\pm$0.28    &   & 25& 6& & & & && &A &     \\
\noalign{\smallskip}
 16 & 0036.5-7225 & 00 36 30.5         & -72 25 40                          & 17          & 2.50$\pm$0.87                               & 0.86$\pm$0.41  & -0.29$\pm$0.29   &   & 20&12& & & & && &A & quasar\\
 17 & 0036.6-7227 & 00 36 39.8         & -72 27 41                          & 17          & 4.23$\pm$1.01                               & 1.0            & 0.42$\pm$0.22    &   & 33&14& & & & && &A & quasar\\
 18 & 0036.9-7339 & 00 36 59.9         & -73 39 43                          & 46          & 5.12$\pm$1.63                               & 0.28$\pm$0.36  & -0.31$\pm$0.31   &   & 12&25& & & & && &   &     \\
 19 & 0036.9-7138 & 00 36 59.9         & -71 38 07                          & 31          & 24.6$\pm$2.8                                & 1.0            & 0.38$\pm$0.11    &   & 83&36& & & & && &   &rad  \\
 20 & 0037.3-7214 & 00 37 19.5         & -72 14 08                          & 10          & 503$\pm$10                                  & -0.97$\pm$0.01 & -0.95$\pm$0.06   & 3 &10410&0& & & &&&Sl& &rad? \\
\noalign{\smallskip}
 21 & 0037.3-7217 & 00 37 19.5         & -72 17 58                          & 12          & 6.92$\pm$1.35                              & 0.65$\pm$0.20  & 0.04$\pm$0.19     &   & 81& 4& & & & && &A &quasar\\
 22 & 0037.5-7224 & 00 37 33.9         & -72 24 36                          & 19          & 3.19$\pm$1.06                              & 0.51$\pm$0.37  & -0.48$\pm$0.29    &   & 18&10& & & & && &   &     \\
 23 & 0037.6-7229 & 00 37 38.5         & -72 29 34                          & 15          & 5.32$\pm$1.14                              & 1.0            & 0.13$\pm$0.21     &   & 45&15& & & & && &   &     \\
 24 & 0038.0-7344 & 00 38 01.6         & -73 44 38                          & 26          & 1.62$\pm$0.59                              & 1.0             & 1.0              &   & 14&22& & & & &&Bw& &     \\
 25 & 0038.0-7327 & 00 38 02.4         & -73 27 43                          & 32          & 4.73$\pm$1.41                              & 0.09$\pm$0.31  & 0.27$\pm$0.34     &   & 23&23& & & & &&F & &     \\
\noalign{\smallskip}
 26 & 0038.6-7310 & 00 38 36.3         & -73 10 22                          & 26          & 31.7$\pm$2.5                               & 1.0            & 0.39$\pm$0.07     & 6 &167&34& & & & &&Bl&A &rad \\
 27 & 0038.7-7214 & 00 38 47.2         & -72 14 08                          & 16          & 2.53$\pm$0.96                              & 0.64$\pm$0.46  & 0.57$\pm$0.29     & &20&7& & & & &&D &   &      \\
 28 & 0038.8-7208 & 00 38 48.8         & -72 08 21                          & 16          & 1.81$\pm$0.58                              & 1.0            & 0.21$\pm$0.32     & &18&9& & & & &&  &    &      \\
 29 & 0038.8-7205 & 00 38 52.8         & -72 05 31                          & 13          & 6.71$\pm$1.26                              & 0.78$\pm$0.19  & 0.29$\pm$0.17     & &80&11& & & & &&  &    &      \\
 30 & 0039.2-7340 & 00 39 15.8         & -73 40 50                          & 17          & 2.42$\pm$0.63                              & 1.0            & 0.57$\pm$0.24     & &25&16& & & & &&Bw&AB  &      \\
\noalign{\smallskip}
 31 & 0039.4-7330 & 00 39 27.0         & -73 30 56                          & 21          & 1.17$\pm$0.51                              & 1.0            & -0.37$\pm$0.30    & &17&16& & & & &&A &    &      \\
 32 & 0039.5-7153 & 00 39 35.5         & -71 53 03                          & 27          & 5.14$\pm$1.52                              & 0.77$\pm$0.37  & 0.16$\pm$0.25     & &23&24& & & & &&  &   &      \\
 33 & 0040.0-7345 & 00 40 01.0         & -73 45 45                          & 13          & 15.4$\pm$1.54                              & -0.04$\pm$0.10 & -0.07$\pm$0.13    & &146&14& & & F3V & 8.9 & 3 &F &   &star  \\
 34 & 0041.0-7336 & 00 41 05.7         & -73 36 38                          & 16          & 9.65$\pm$1.33                              & 0.89$\pm$0.15  & -0.14$\pm$0.12    & 30 &67&8& & & & &&R &   &rad \\
 35 & 0041.7-7326 & 00 41 42.3         & -73 26 19                          & 17          & 2.46$\pm$0.84                              & 0.97$\pm$0.46  & 0.67$\pm$0.23     & &30&13& & & & &&Bw&AB  &      \\
\noalign{\smallskip}
 36 & 0041.7-7222 & 00 41 47.5         & -72 22 08                          & 35          & 6.30$\pm$1.72                              & 0.91$\pm$0.35  & 0.27$\pm$0.22     & &23&22& & & & &&Sw/D&     &      \\
 37 & 0041.9-7308 & 00 41 58.5         & -73 08 01                          & 60          & 5.85$\pm$1.28                              & 1.0            & 0.59$\pm$0.26     & &15&31& & & & &&Bw&AB  &rad \\
 38 & 0042.2-7338 & 00 42 12.1         & -73 38 04                          & 111         & 5.79$\pm$14.8                              & -1.0           &                   & 214 &24&44& & & & &&Sw/D& &      \\
 39 & 0042.6-7340 & 00 42 41.3         & -73 40 37                          & 14          & 2.59$\pm$0.80                              & 0.96$\pm$0.41  & 0.60$\pm$0.23     &  &37&2& & & & &&Bw&AB  &      \\
 40 & 0043.3-7335 & 00 43 23.2         & -73 35 17                          & 15          & 0.93$\pm$0.33                              & 1.0            & 0.88$\pm$0.22     & &18&4& & & & &&Bw&  &      \\
\noalign{\smallskip}
 41 & 0043.7-7355 & 00 43 47.6         & -73 55 21                          & 22          & 2.67$\pm$0.84                              & 0.42$\pm$0.37  & 0.30$\pm$0.27     & &19&17& & & & &&  &    &      \\
 42 & 0043.9-7342 & 00 43 55.9         & -73 42 09                          & 14          & 1.46$\pm$0.45                              & 1.0            & 0.19$\pm$0.31     & &22&6& & & & && &A   &      \\
 43 & 0043.9-7322 & 00 43 57.7         & -73 22 24                          & 18          & 4.53$\pm$1.01                              & 0.67$\pm$0.26  & 0.84$\pm$0.14     & &46&17& & & & &&  &    &      \\
 44 & 0044.0-7415 & 00 44 02.5         & -74 15 53                          & 39          & 17.3$\pm$3.05                              & 0.09$\pm$0.18  & 0.01$\pm$0.19     & &32&38& & & F6IV & 8.0 & 20 &F &   &star  \\
 45 & 0044.2-7350 & 00 44 14.9         & -73 50 09                          & 21          & 1.03$\pm$0.39                              & 1.00$\pm$0.00  & 0.05$\pm$0.38     & &10&13& & & F0V & 9.2 & 3 & &A     &      \\
\noalign{\smallskip}
 46 & 0044.3-7336 & 00 44 21.9         & -73 36 34                          & 18          & 1.20$\pm$0.39                              & 1.0            & 0.81$\pm$0.23     & &16&6& & & & &&Bw&  &      \\
 47 & 0044.3-7330 & 00 44 23.2         & -73 30 37                          & 16          & 2.51$\pm$0.73                              & 0.85$\pm$0.36  & 0.43$\pm$0.24     & &34&10& & & F0V & 11.3 & 47 & &A    &      \\
 48 & 0045.1-7303 & 00 45 08.5         & -73 03 56                          & 24          & 2.92$\pm$0.54                              & 1.0            & -0.19$\pm$0.19    & &30&26& & & & &&A &H   &      \\
 49 & 0045.1-7341 & 00 45 08.6         & -73 41 57                          & 18          & 2.31$\pm$0.79                              & -0.04$\pm$0.34 & 0.78$\pm$0.40     & &11&10& & & & &&F &   &rad \\
 50 & 0045.4-7328 & 00 45 27.1         & -73 28 36                          & 58          & 33.6$\pm$7.3                               & -0.43$\pm$0.19 & -0.46$\pm$0.75    & &23&27& & & & &&F &   &      \\
            \noalign{\smallskip}
            \hline
            \noalign{\smallskip}
\end{longtable}
\noindent
\clearpage
\vfill \eject

  \begin{longtable}[l]{rccccccccrrrcrrcclll}
   \hline 
  \noalign{\smallskip}
(1) &(2)          &(3)                 &(4)                                 &(5)          &(6)                                         &(7)             &(8)                &(9)         &(10)   &(11)         &(12)  &(13)                   &(14)  &(15)   &(16)         &(17)   &(18) &(19) \\
    & Source      & RA                 & Dec                                &             & \multicolumn{1}{c}{Count Rate}              & \multicolumn{2}{c}{Hardness Ratio} &             &       &             & \multicolumn{2}{c}{Einstein} & \multicolumn{3}{c}{Simbad} &       &     \\
 No.& Name        & (J2000)            & (J2000)                            & $\rm P_{\rm e}$ & Total                                       & HR1            & HR2               & Ext         & LH    & $\Delta$    & ID   &  Dist                 & Type & Mag   & Dist        & Class & Class & Note\\
    & RX~J        & \p0h~\p0m~\p0s\p0~ & \p0\D\p0\p0$\arcmin$~\p0$\arcsec$  & ($\arcsec$) & \multicolumn{1}{c}{$\rm (10^{-3}\ s^{-1})$} &                &                   & ($\arcsec$) &       & ($\arcmin$) &      & ($\arcsec$)           &      &       &($\arcsec$)  & &(refined)      &     \\
            \noalign{\smallskip}
            \hline
            \noalign{\smallskip}
 51 & 0045.6-7313 & 00 45 37.1         & -73 13 49                          & 50          & 4.19$\pm$1.73                              & 0.79$\pm$0.58  & 0.13$\pm$0.37     & &10&27& & & B2 & 12.9 & 11 & &      &rad snr? \\
 52 & 0045.6-7335 & 00 45 38.2         & -73 35 27                          & 17          & 1.89$\pm$0.51                              & 1.0            & 0.46$\pm$0.25     & &25&12& & & & && &AB    &      \\
 53 & 0045.6-7352 & 00 45 41.9         & -73 52 57                          & 20          & 3.08$\pm$0.74                              & 1.0            & 0.69$\pm$0.22     & &30&19& & & & &&Bw&  &      \\
 54 & 0045.8-7340 & 00 45 50.6         & -73 40 15                          & 15          & 3.67$\pm$0.69                              & 1.0            & 0.31$\pm$0.18     & &52&12& & & B5I & 12.6 & 18 & &A    &      \\
 55 & 0046.5-7308 & 00 46 31.8         & -73 08 23                          & 14          & 10.7$\pm$1.00                              & 0.78$\pm$0.10  & -0.28$\pm$0.08    & 31 &196&19& 15 & 35 & & &  &R &   &      \\
\noalign{\smallskip}
 56 & 0046.5-7300 & 00 46 35.3         & -73 00 59                          & 25          & 4.79$\pm$1.02                              & -0.49$\pm$0.18 & 0.39$\pm$0.42     & 47 &52&22& & & & &&F &   &      \\
 57 & 0047.2-7239 & 00 47 17.2         & -72 39 39                          & 51          & 2.11$\pm$0.71                              & 1.0            & 1.0               & &11&33& 19 & 60 & & &  &Bw&   &rad \\
 58 & 0047.3-7312 & 00 47 21.3         & -73 12 18                          & 11          & 11.6$\pm$0.84                              & 1.0            & 0.49$\pm$0.06     & &481&15& 18 & 46 & & & 82 &Bw&  &      \\
 59 & 0047.4-7305 & 00 47 27.1         & -73 05 59                          & 23          & 2.28$\pm$0.45                              & 1.0            & -0.18$\pm$0.18    & &19&16& 0 & & M & 12.9 & 70 &A &   &      \\
 60 & 0047.5-7308 & 00 47 30.8         & -73 08 46                          & 18          & 35.1$\pm$2.0                               & 1.0            & 0.21$\pm$0.06     & &361&38& 16 & 34 & B0 & 12.5 & 88 &A &    &      \\
\noalign{\smallskip}
 61 & 0047.6-7309 & 00 47 38.1         & -73 09 21                          & 11          & 22.92$\pm$1.21                              & 0.92$\pm$0.04  & 0.21$\pm$0.05     & 20 &837&14& 16 & 74 & B5I & 12.2 & 83 &R &   &rad \\
 62 & 0048.2-7331 & 00 48 16.3         & -73 31 44                          & 10          & 187$\pm$3                                  & -0.97$\pm$0.01 & -0.92$\pm$0.12    & 15 &8852&21& & & & & 73 &Sl&  &      \\
 63 & 0048.3-7319 & 00 48 21.6         & -73 19 19                          & 15          & 3.87$\pm$0.52                              & 1.0            & 0.40$\pm$0.13     & 21 &57&12& 21 & 33 & & & 14 & &      &rad snr\\
 64 & 0048.4-7308 & 00 48 24.8         & -73 08 39                          & 20          & 0.78$\pm$0.38                              & 0.89$\pm$0.73  & -0.24$\pm$0.32    & &12&11& & & & &&   &   &      \\
 65 & 0048.5-7323 & 00 48 30.8         & -73 23 35                          & 15          & 1.54$\pm$0.34                              & 1.0            & 0.84$\pm$0.17     & &38&14& & & & &&Bw&  &      \\
\noalign{\smallskip}
 66 & 0048.5-7302 & 00 48 32.4         & -73 02 18                          & 13          & 5.30$\pm$0.59                              & 1.0            & 0.47$\pm$0.10     & &137&15& & & A0I & 12.8 & 79 &Bw&  &      \\
 67 & 0048.9-7306 & 00 48 57.6         & -73 06 07                          & 18          & 1.61$\pm$0.35                              & 1.0            & -0.21$\pm$0.22    & &22&11& & & & & 37 &A &H   &rad \\
 68 & 0049.0-7314 & 00 49 05.9         & -73 14 06                          & 11          & 12.5$\pm$0.84                              & 1.0            & 0.11$\pm$0.06     & 15 &450&7& 22 & 2 & & 16.3 & 83 &R &   &rad \\
 69 & 0049.0-7250 & 00 49 05.9         & -72 50 55                          & 22          & 45.2$\pm$7.0                               & 1.0            & 0.86$\pm$0.10     & &64&24& & & & & 75 &Bl&  &      \\
 70 & 0049.4-7310 & 00 49 27.6         & -73 10 53                          & 12          & 2.32$\pm$0.48                              & 0.86$\pm$0.25  & 0.86$\pm$0.11     & &85&6& & & B1 & 13.7 & 85 &Bw&  &      \\
\noalign{\smallskip}
 71 & 0049.8-7324 & 00 49 50.0         & -73 24 57                          & 18          & 1.40$\pm$0.45                              & 0.81$\pm$0.43  & 0.68$\pm$0.23     & &22&12& & & & & 55 & &AB     &      \\
 72 & 0050.6-7315 & 00 50 41.9         & -73 15 56                          & 11          & 3.35$\pm$0.44                              & 1.0            & 0.57$\pm$0.11     & &132&2& & & & &&Bw&A &      \\
 73 & 0050.7-7226 & 00 50 43.1         & -72 26 37                          & 53          &                                            &                &                   & 93 &31&35& & & & &&D &   &      \\
 74 & 0050.7-7332 & 00 50 45.9         & -73 32 35                          & 28          & 1.25$\pm$0.37                              & 1.0            &-0.32$\pm$0.29     & &14&19& & & & &&A &    &      \\
 75 & 0050.8-7341 & 00 50 48.9         & -73 41 07                          & 42          & 4.77$\pm$1.45                              & 1.0            & 0.40$\pm$0.31     & 70 &43&33& & & & && &AB     &rad \\
\noalign{\smallskip}
 76 & 0050.9-7310 & 00 50 54.3         & -73 10 07                          & 11          & 2.67$\pm$0.46                              & 0.99$\pm$0.20  & 0.49$\pm$0.13     & &106&4&&& & &&Bw&A &      \\
 77 & 0051.0-7321 & 00 51 03.9         & -73 21 24                          & 10          & 118$\pm$2.2                                & 0.79$\pm$0.02  &-0.26$\pm$0.02     & 28 &3973&8& 24 & 17 & & & 13 &R &   &rad \\
 78 & 0051.3-7250 & 00 51 18.7         & -72 50 36                          & 18          & 1.52$\pm$0.51                              & 0.28$\pm$0.38  & 0.94$\pm$0.30     & &17&15& & & B1 & 13.6 & 61 &   &   &      \\
 79 & 0051.3-7216 & 00 51 22.3         & -72 16 37                          & 30          & 5.66$\pm$0.79                              & 1.0            & 0.65$\pm$0.14     & &52&31& & & B2I & 13.5 & 86 &Bw&  &      \\
 80 & 0051.6-7304 & 00 51 40.6         & -73 04 09                          & 18          & 0.77$\pm$0.38                              & 0.91$\pm$0.75  & 1.0               & &18&10& & & & &&   &   &      \\
\noalign{\smallskip}
 81 & 0051.7-7341 & 00 51 42.7         & -73 41 51                          & 67          & 1.64$\pm$0.68                              & 1.0            & 1.0               & &10&28& & & & &&Bw&  &      \\
 82 & 0051.8-7310 & 00 51 49.2         & -73 10 30                          & 10          & 12.3$\pm$0.86                              & 0.98$\pm$0.04  & 0.33$\pm$0.06     & &744&6& 25 & 34 & & 15.2 & 68 &  &    &      \\
 83 & 0051.8-7231 & 00 51 53.0         & -72 31 44                          & 10          & 102$\pm$2                                   & 0.95$\pm$0.01  & 0.47$\pm$0.02     & 10 &4489&18& 27 & 14 & B2 & 14.0 & 81 &Bl&  &      \\
 84 & 0052.1-7319 & 00 52 11.2         & -73 19 13                          & 10          & 18.5$\pm$1.0                               & 0.95$\pm$0.04  & 0.61$\pm$0.04     & &1162&8& 29 & 27 & O9 & 12.5 & 56 &Bl&A &      \\
 85 & 0052.2-7301 & 00 52 16.9         & -73 01 54                          & 23          & 1.67$\pm$0.54                              & 0.09$\pm$0.34  & 0.81$\pm$0.37     & &12&14& & & & &&F &   &      \\
\noalign{\smallskip}
 86 & 0052.5-7237 & 00 52 30.0         & -72 37 18                          & 16          & 6.41$\pm$0.90                              & 0.90$\pm$0.16  & -0.49$\pm$0.11    & 26 &76&12& 30 & 50 & BN & 13.6 & 82 &R &   &rad snr \\
 87 & 0052.6-7247 & 00 52 40.8         & -72 47 14                          & 16          & 0.95$\pm$0.28                              & 1.0            & 0.28$\pm$0.30     & &18&8& 31 & 85 & B0 & 14.2 & 73 & &A   &      \\
 88 & 0052.8-7259 & 00 52 51.7         & -72 59 56                          & 21          & 2.18$\pm$0.67                              & 0.98$\pm$0.45  & 0.02$\pm$0.22     & &22&16& & & & &&A &    &      \\
 89 & 0053.0-7158 & 00 53 02.1         & -71 58 05                          & 27          & 4.54$\pm$0.69                              & 1.0            & 0.80$\pm$0.16     & &55&30& 32 & 44 & & &  &Bw&  &      \\
 90 & 0053.0-7239 & 00 53 05.6         & -72 39 14                          & 22          & 0.96$\pm$0.46                              & 0.87$\pm$0.71  & -0.50$\pm$0.31    & &10&9& & & B6I & 10.8 & 74 & &    &rad snr\\
\noalign{\smallskip}
 91 & 0053.1-7311 & 00 53 06.2         & -73 11 52                          & 17          & 0.78$\pm$0.37                              & 0.46$\pm$0.63  & 0.76$\pm$0.35     & &11&10& & & B2 & 14.0 & 88 & &F   &      \\
 92 & 0053.1-7337 & 00 53 06.9         & -73 37 21                          & 48          & 2.50$\pm$0.62                              & 1.0$\pm$0.0    & 0.20$\pm$0.26     & &11&26& & & & &&- &A     &      \\
 93 & 0053.3-7236 & 00 53 18.2         & -72 36 05                          & 27          & 4.44$\pm$0.95                              & 0.80$\pm$0.27  & -0.33$\pm$0.17    & &20&11& & & & &&A &    &      \\
 94 & 0053.5-7227 & 00 53 32.2         & -72 27 04                          & 23          & 2.62$\pm$0.67                              & 1.0            & 0.35$\pm$0.26     & 49 &76&19& & & & && &A    &      \\
 95 & 0053.6-7201 & 00 53 41.9         & -72 01 02                          & 49          &                                             &                &                   & 133 &43&34& & & & &&R/D&  &      \\
\noalign{\smallskip}
 96 & 0053.8-7129 & 00 53 48.7         & -71 29 26                          & 33          & 2.68$\pm$0.81                              & 1.0            & 0.22$\pm$0.29     & &14&23& & & & && &AB     &      \\
 97 & 0053.8-7252 & 00 53 53.2         & -72 52 19                          & 17          & 0.54$\pm$0.21                              & 1.0            & 0.91$\pm$0.39     & &11&7& & & & & 64 &Bw&AB  &      \\
 98 & 0053.9-7226 & 00 53 57.3         & -72 26 35                          & 12          & 11.4$\pm$0.78                              & 1.0            & 0.58$\pm$0.06     & 19 &380&22& 34 & 15 & & 14.4 & 37 &Bw/D& &      \\
 99 & 0054.3-7330 & 00 54 19.0         & -73 30 31                          & 30          & 0.91$\pm$0.34                              & 1.0            & 0.29$\pm$0.40     & &11&23& & & & && &AB     &      \\
 100 & 0054.5-7340 & 00 54 31.7        & -73 40 56                          & 14          & 372$\pm$20                                 & 0.97$\pm$0.02  & 0.63$\pm$0.04     & 34 &907&32& & & & & 18 &Bl&  &      \\
            \noalign{\smallskip}
            \hline
            \noalign{\smallskip}
\end{longtable}
\noindent
\clearpage
\vfill \eject

  \begin{longtable}[l]{rccccccccrrrcrrcclll}
   \hline 
  \noalign{\smallskip}
(1) &(2)          &(3)                 &(4)                                 &(5)          &(6)                                        &(7)             &(8)                &(9)         &(10)   &(11)         &(12)  &(13)                   &(14)  &(15)   &(16)         &(17)   &(18) &(19) \\
    & Source      & RA                 & Dec                                &             & \multicolumn{1}{c}{Count Rate}              & \multicolumn{2}{c}{Hardness Ratio} &             &       &             & \multicolumn{2}{c}{Einstein} & \multicolumn{3}{c}{Simbad} &       &     \\
 No.& Name        & (J2000)            & (J2000)                            & $\rm P_{\rm e}$ & Total                                        & HR1            & HR2               & Ext         & LH    & $\Delta$    & ID   &  Dist                 & Type & Mag   & Dist        & Class & Class & Note\\
    & RX~J        & \p0h~\p0m~\p0s\p0~ & \p0\D\p0\p0$\arcmin$~\p0$\arcsec$  & ($\arcsec$) & \multicolumn{1}{c}{$\rm (10^{-3}\ s^{-1})$} &                &                   & ($\arcsec$) &       & ($\arcmin$) &      & ($\arcsec$)           &      &       &($\arcsec$)  & &(refined)      &     \\
            \noalign{\smallskip}
            \hline
            \noalign{\smallskip}
 101 & 0054.5-7218 & 00 54 35.1        & -72 18 07                          & 17          & 3.41$\pm$0.60                              & 0.91$\pm$0.22  & 0.62$\pm$0.13     & &64&17& & & B0 & 14.2 & 82 &  Bw &  &      \\
 102 & 0054.9-7245 & 00 54 54.7        & -72 45 02                          & 11          & 3.12$\pm$0.58                              & 0.64$\pm$0.22  & 0.60$\pm$0.14     & &85&2& & & B1II & 14.0 & 76 &   &   &      \\
 103 & 0054.9-7226 & 00 54 57.3        & -72 26 39                          & 11          & 27.3$\pm$1.15                              & 1.0            & 0.55$\pm$0.03     & 16 &1304&18& 35 & 13 & & & 41 &  Bl &  &      \\
 104 & 0055.2-7238 & 00 55 17.7        & -72 38 53                          & 12          & 3.32$\pm$0.58                              & 0.79$\pm$0.20  & 0.77$\pm$0.12     & &97&8& & & & &&  Bw &  &      \\
 105 & 0055.4-7210 & 00 55 29.2        & -72 10 53                          & 10          & 24.7$\pm$1.04                              & 1.0            & 0.36$\pm$0.04     & 8 &1632&14& 36 & 12 & & &  &  Bl & A &rad \\
\noalign{\smallskip}
 106 & 0055.6-7228 & 00 55 36.8        & -72 28 26                          & 14          & 4.36$\pm$0.51                              & 1.0            & 0.33$\pm$0.11     & &88&17& & & & &&  & A     &      \\
 107 & 0055.6-7234 & 00 55 39.9        & -72 34 55                          & 28          & 1.12$\pm$0.31                              & 1.0            & 0.49$\pm$0.25     & &12&21& & & & && D & &          \\
 108 & 0055.6-7116 & 00 55 41.6        & -71 16 58                          & 40          & 5.24$\pm$1.41                              & 1.0            & 0.18$\pm$0.30     & &14&23& & & & &&  & A    &      \\
 109 & 0055.7-7138 & 00 55 47.0        & -71 38 11                          & 19          & 4.22$\pm$1.07                              & -0.42$\pm$0.21 & -0.20$\pm$0.40    & &20&13& & & F5/F6 & 11.5 & 12 &  F &   &star  \\
 110 & 0055.7-7331 & 00 55 47.7        & -73 31 04                          & 17          & 12.7$\pm$1.0                               & 1.0            & 0.34$\pm$0.08     & &176&28& & & & && & A     &      \\
\noalign{\smallskip}
 111 & 0055.8-7241 & 00 55 51.7        & -72 41 58                          & 13          & 2.45$\pm$0.50                              & 0.63$\pm$0.23  & -0.01$\pm$0.19    & &53&7& & & & &&   &   &      \\
 112 & 0055.9-7257 & 00 55 58.1        & -72 57 10                          & 70          & 22.0$\pm$6.7                               & -1.0           &                   & 91 &17&28& & & & && Sl/D & &      \\
 113 & 0056.4-7159 & 00 56 26.9        & -71 59 47                          & 24          &                                            &                &                   & &33&19& & & & &&  D &  &      \\
 114 & 0056.6-7220 & 00 56 41.7        & -72 20 24                          & 12          & 1.95$\pm$0.39                              & 0.86$\pm$0.25  & 0.55$\pm$0.15     & &64&8& & & & &&  Bw &  &rad snr?\\
 115 & 0056.8-7310 & 00 56 49.3        & -73 10 41                          & 41          & 3.19$\pm$0.66                              & 1.0            & 0.37$\pm$0.23     & &15&26& & & & && & A    &      \\
\noalign{\smallskip}
 116 & 0056.9-7211 & 00 56 54.1        & -72 11 57                          & 14          & 0.73$\pm$0.20                              & 1.0            & 0.59$\pm$0.26     & &19&8& & & & &&  Bw & AB  &      \\
 117 & 0057.0-7131 & 00 57 03.6        & -71 31 58                          & 19          & 1.85$\pm$0.67                              & 1.0            & 0.46$\pm$0.37     & &11&8& & & & &&  & AB    &      \\
 118 & 0057.2-7145 & 00 57 14.6        & -71 45 23                          & 20          & 1.48$\pm$0.59                              & 1.0            & -0.26$\pm$0.39    & &11&11& & & & && A &    &      \\
 119 & 0057.2-7156 & 00 57 17.8        & -71 56 20                          & 52          & 7.51$\pm$1.00                              & -0.52$\pm$0.12 & -0.46$\pm$0.31    & &13&21& & & & &&  F &   &      \\
 120 & 0057.3-7225 & 00 57 20.5        & -72 25 26                          & 11          & 7.48$\pm$0.62                              & 0.83$\pm$0.07  & 0.64$\pm$0.06     & &403&9& 39 & 20 & & &  &  Bw &  &      \\
\noalign{\smallskip}
 121 & 0057.3-7325 & 00 57 21.9        & -73 25 08                          & 25          & 7.96$\pm$0.95                              & 1.0            & 0.69$\pm$0.11     & &85&31& & & & & 86 &  Bw &  &      \\
 122 & 0057.5-7313 & 00 57 32.4        & -73 13 15                          & 33          & 3.54$\pm$0.68                              & 1.0            & 0.53$\pm$0.22     & &26&29& & & & &&  Bw & AB  &      \\
 123 & 0057.5-7212 & 00 57 33.7        & -72 12 59                          & 11          & 4.64$\pm$0.45                              & 1.0            & 0.44$\pm$0.09     & &255&5&40&18& BE & 13.4 & 37 & Bw & A &    \\
 124 & 0057.8-7202 & 00 57 51.0        & -72 02 40                          & 13          & 2.06$\pm$0.34                              & 1.0            & 0.82$\pm$0.12     & &71&14& 41 & 49 & & &  &  Bw &   &      \\
 125 & 0057.8-7207 & 00 57 51.7        & -72 07 56                          & 11          & 2.92$\pm$0.36                              & 1.0            & 0.58$\pm$0.11     & &129&9& & & & &&  Bw &   &      \\
\noalign{\smallskip}
 126 & 0057.9-7156 & 00 57 58.2        & -71 56 26                          & 26          & 1.98$\pm$0.57                              & 0.88$\pm$0.40  & 0.63$\pm$0.21     & &25&20& & & & &&  D &  &      \\
 127 & 0058.2-7116 & 00 58 17.4        & -71 16 47                          & 27          & 3.28$\pm$0.83                              & 1.0            & 0.21$\pm$0.26     & &19&19& & & & && & A    &      \\
 128 & 0058.3-7218 & 00 58 18.0        & -72 18 03                          & 12          & 7.86$\pm$0.68                              & 0.90$\pm$0.08  & 0.44$\pm$0.08     & 17 &246&1& 42 & 26 & F2IA & 12.3 & 70 & R &   &rad snr      \\
 129 & 0058.3-7229 & 00 58 19.4        & -72 29 52                          & 19          & 0.57$\pm$0.20                              & 1.0            & 0.64$\pm$0.32     & &10&13& & & & &&  Bw & AB  &      \\
 130 & 0058.3-7200 & 00 58 23.6        & -72 00 26                          & 30          & 2.14$\pm$0.45                              & 1.0            & 0.50$\pm$0.21     & &18&16& & & & &&  & AB    &      \\
\noalign{\smallskip}
 131 & 0058.5-7208 & 00 58 30.1        & -72 08 46                          & 15          & 0.65$\pm$0.19                              & 1.0            & 0.33$\pm$0.30     & &13&8& & & & & 28 & & A    &      \\
 132 & 0058.5-7249 & 00 58 34.0        & -72 49 47                          & 70          &                                             &                &                   & 175 &36&42& & & & && R/D &   &      \\
 133 & 0058.5-7146 & 00 58 35.8        & -71 46 01                          & 13          & 25.5$\pm$2.5                               &-0.99$\pm$0.024 & -1.0              & &166&10& & & B0 & 13.2 & 83 &  Sl & &      \\
 134 & 0058.6-7203 & 00 58 37.0        & -72 03 11                          & 20          & 0.80$\pm$0.24                              & 1.0            & 0.20$\pm$0.30     & &11&14& & & & && & A    &      \\
 135 & 0058.6-7135 & 00 58 37.1        & -71 35 56                          & 10          & 355$\pm$7                                  & -0.99$\pm$0.01 & -1.0              & 12 &9230&0& 43 & 3 & & 16.6 & 7 &  Sl &  &      \\
\noalign{\smallskip}
 136 & 0058.9-7255 & 00 58 59.3        & -72 55 50                          & 29          &                                            &                &                   & 120 &184&38& & & & && R/D &   &      \\
 137 & 0059.0-7119 & 00 59 02.4        & -71 19 41                          & 24          & 1.54$\pm$0.56                              & 1.0            & 0.09$\pm$0.37     & &10&16& & & & && & A   &      \\
 138 & 0059.1-7216 & 00 59 08.7        & -72 16 27                          & 18          & 1.34$\pm$0.36                              & -0.16$\pm$0.26 & -0.39$\pm$0.34    & &14&4& & & O7III & 14.1 & 52 &  F &    &star  \\
 139 & 0059.3-7223 & 00 59 21.1        & -72 23 12                          & 10          & 8.07$\pm$0.42                              & 0.86$\pm$0.07  & 0.52$\pm$0.06     & &470&8& & & & &&  Bw &   &      \\
 140 & 0059.5-7210 & 00 59 31.0        & -72 10 09                          & 10          & 33.4$\pm$1.20                              & 0.88$\pm$0.03  & -0.08$\pm$0.04    & 22 &1232&9& 44 & 24 & O9 & & 74 &  R &    &rad \\
\noalign{\smallskip}
 141 & 0059.6-7138 & 00 59 40.8        & -71 38 05                          & 23          & 4.05$\pm$1.17                              & -1.0           &                   &    &12&6& & & & 17.4 & 8 & Sw & &      \\
 142 & 0059.7-7148 & 00 59 44.0        & -71 48 17                          & 20          & 1.06$\pm$0.45                              & 1.0            & 1.0               & &12&13& & & & &&  Bw &   &      \\
 143 & 0100.0-7157 & 01 00 05.3        & -71 57 21                          & 22          & 2.41$\pm$0.86                              & 1.0            & 0.58$\pm$0.36     & &11&10& & & B5I & 13.2 & 54 & & AB    &      \\
 144 & 0100.1-7118 & 01 00 07.6        & -71 18 03                          & 21          & 2.67$\pm$0.72                              & 1.0            & 0.30$\pm$0.26     & &23&19& & & & && & A    &      \\
 145 & 0100.2-7307 & 01 00 12.6        & -73 07 32                          & 28          & 12.3$\pm$1.59                              & 0.02$\pm$0.13  & -0.01$\pm$0.15    & &50&33& & & & &  &  F &    &      \\
\noalign{\smallskip}
 146 & 0100.2-7220 & 01 00 12.8        & -72 20 08                          & 14          & 1.10$\pm$0.34                              & 0.94$\pm$0.45  & 0.73$\pm$0.20     & &31&10& & & & &  &  &    &      \\
 147 & 0100.2-7204 & 01 00 15.1        & -72 04 40                          & 19          & 1.40$\pm$0.33                              & 1.0            & 0.77$\pm$0.21     & &24&15& & & & &&  Bw &   &      \\
 148 & 0100.3-7133 & 01 00 18.9        & -71 33 21                          & 24          & 7.91$\pm$1.20                              & 1.0            & -0.33$\pm$0.15    & 29 &37&9& & & O9 & & 25 &  A &    &rad snr\\
 149 & 0100.3-7241 & 01 00 21.3        & -72 41 27                          & 44          & 2.01$\pm$0.58                              & 1.0            & 0.43$\pm$0.32     & &10&26& & & O9III & 13.5 & 31 & & AB     &      \\
 150 & 0100.4-7201 & 01 00 24.3        & -72 01 23                          & 30          & 4.18$\pm$0.63                              & 1.0            & -0.32$\pm$0.14    & 47 &14&18& & & A0I & 12.8 & 88 & A &    &      \\
            \noalign{\smallskip}
            \hline
            \noalign{\smallskip}
\end{longtable}
\noindent
\clearpage
\vfill \eject

  \begin{longtable}[l]{rcccccccrrrcrrcclll}
   \hline 
  \noalign{\smallskip}
(1) &(2)          &(3)                 &(4)                                 &(5)          &(6)                                      &(7)             &(8)                &(9)         &(10)   &(11)         &(12)  &(13)                   &(14)  &(15)   &(16)         &(17)   &(18) &(19) \\
    & Source      & RA                 & Dec                                &             & \multicolumn{1}{c}{Count Rate}              & \multicolumn{2}{c}{Hardness Ratio} &             &       &             & \multicolumn{2}{c}{Einstein} & \multicolumn{3}{c}{Simbad} &       &     \\
 No.& Name        & (J2000)            & (J2000)                            & $\rm P_{\rm e}$ & Total                                  & HR1            & HR2               & Ext         & LH    & $\Delta$    & ID   &  Dist                 & Type & Mag   & Dist        & Class & Class & Note\\
    & RX~J        & \p0h~\p0m~\p0s\p0~ & \p0\D\p0\p0$\arcmin$~\p0$\arcsec$  & ($\arcsec$) & \multicolumn{1}{c}{$\rm (10^{-3}\ s^{-1})$} &                &                   & ($\arcsec$) &       & ($\arcmin$) &      & ($\arcsec$)           &      &       &($\arcsec$)  & &(refined)      &     \\
            \noalign{\smallskip}
            \hline
            \noalign{\smallskip}
 151 & 0100.4-7149 & 01 00 26.6        & -71 49 28                          & 15          & 2.64$\pm$1.06                              & 0.95$\pm$0.51  & 0.91$\pm$0.16     & &25&4& & & & &&   &   &      \\
 152 & 0100.5-7259 & 01 00 34.3        & -72 59 55                          & 29          & 13.8$\pm$1.79                              & 0.07$\pm$0.13  & -0.30$\pm$0.14    & &54&30& & & & &&  F &    &      \\
 153 & 0100.7-7211 & 01 00 43.4        & -72 11 34                          & 10          & 23.9$\pm$1.05                              & 0.97$\pm$0.03  & 0.33$\pm$0.04     & 8 &1481&13& 45 & 37 & K5 & & 48 & A & A &rad \\
 154 & 0100.8-7214 & 01 00 48.9        & -72 14 15                          & 17          & 1.00$\pm$0.26                              & 1.0            & 0.56$\pm$0.27     & &16&12& & & B1I & 13.1 & 16 & & AB    &      \\
 155 & 0100.9-7222 & 01 00 57.4        & -72 22 25                          & 14          & 1.80$\pm$0.33                              & 1.0            & 0.86$\pm$0.14     & &52&14& & & & 13.0 & 44 &  Bw &   &      \\
\noalign{\smallskip}
 156 & 0101.0-7151 & 01 01 00.6        & -71 51 53                          & 47          & 2.61$\pm$1.77                              & -0.62$\pm$0.57 & 0.88$\pm$3.80     & &11&20& & & & &&    &  &      \\
 157 & 0101.0-7211 & 01 01 03.2        & -72 11 29                          & 24          & 36.1$\pm$3.6                               & 1.0            & 0.53$\pm$0.09     & 55 &107&22& 45 & 67 & & 14.9 & 25 & Bl/D &  &      \\
 158 & 0101.0-7206 & 01 01 03.5        & -72 06 56                          & 10          & 27.7$\pm$1.18                              & 0.91$\pm$0.03  & 0.45$\pm$0.04     & 11 &1368&16& & & O8V & 14.6 & 38 &  Bl &   &      \\
 159 & 0101.1-7234 & 01 01 06.6        & -72 34 39                          & 24          & 1.46$\pm$0.51                              & 1.0            & -0.08$\pm$0.44    & 45 &45&22& & & G5 & 11.7 & 69 &  &    &      \\
 160 & 0101.3-7211 & 01 01 18.4        & -72 11 22                          & 15          & 29.3$\pm$1.7                               & 1.0            & 0.44$\pm$0.06     & 36 &316&31& & & & 14.9 & 78 & Bl/D &  &      \\
\noalign{\smallskip}
 161 & 0101.3-7118 & 01 01 20.4        & -71 18 14                          & 59          & 5.10$\pm$2.38                              & -0.83$\pm$0.39 & -1.0              & &12&22& & & & 16.9 & 54 &  D &  &      \\
 162 & 0101.6-7204 & 01 01 37.3        & -72 04 23                          & 16          & 4.75$\pm$0.71                              & 0.78$\pm$0.17  & 0.59$\pm$0.12     & &79&20& 46 & 40 & B & 13.9 & 83 & Bw/D &  &      \\
 163 & 0101.6-7126 & 01 01 38.3        & -71 26 46                          & 26          & 1.59$\pm$0.84                              & 0.69$\pm$0.71  & 0.36$\pm$0.35     & &10&17& & & & && & AB     &      \\
 164 & 0101.6-7154 & 01 01 40.4        & -71 54 26                          & 15          & 6.78$\pm$1.51                              & 0.64$\pm$0.23  & 0.03$\pm$0.22     & &42&6& 79 & 41 & B0 & 14.3 & 85 &      &      \\
 165 & 0101.8-7249 & 01 01 51.3        & -72 49 03                          & 29          & 16.7$\pm$1.8                               & -0.57$\pm$0.09 & 0.12$\pm$0.27     & 88 &49&24& & & & &&  R &    &      \\
\noalign{\smallskip}
 166 & 0101.8-7223 & 01 01 52.2        & -72 23 26                          & 13          & 7.79$\pm$0.77                              & 0.94$\pm$0.10  & 0.73$\pm$0.07     & 17 &221&18& 47 & 65 & & & 48 &  Bw &   &      \\
 167 & 0101.8-7233 & 01 01 52.9        & -72 33 18                          & 41          & 1.99$\pm$0.53                              & 1.0            & 884$\pm$339       & &14&23& & & O9 & 13.8 & 51 &  Bw &   &      \\
 168 & 0102.1-7236 & 01 02 11.3        & -72 36 49                          & 106         &                                            &                &                   & &12&36& & & & &&  D   &      \\
 169 & 0102.5-7239 & 01 02 31.5        & -72 39 40                          & 20          & 1.70$\pm$0.40                              & 1.0            & 0.20$\pm$0.24     & &23&16& & & B0.5V & 14.7 & 52 & & A     &      \\
 170 & 0102.6-7232 & 01 02 41.5        & -72 32 36                          & 12          & 8.66$\pm$0.98                              & 0.72$\pm$0.11  & 0.62$\pm$0.09     & &178&15& & & K0 & & 86 &  Bw &   &      \\
\noalign{\smallskip}
 171 & 0102.7-7214 & 01 02 44.8        & -72 14 34                          & 29          & 0.79$\pm$0.59                              & 1.0            & 0.96$\pm$1.23     & &37&24& & & & && & AB     &      \\
 172 & 0102.8-7216 & 01 02 50.0        & -72 16 24                          & 36          & 1.88$\pm$0.55                              & 1.0            & 0.56$\pm$0.39     & &20&21& & & B2I & 12.3 & 76 &  D &  &      \\
 173 & 0102.9-7111 & 01 02 58.6        & -71 11 17                          & 73          & 17.3$\pm$4.9                               & -1.0           &                   & 127 &20&32& & & & && Sl/D & &      \\
 174 & 0103.0-7225 & 01 03 05.0        & -72 25 11                          & 21          & 1.58$\pm$0.43                              & 1.0            & 0.62$\pm$0.27     & &17&15& & & B & 13.7 & 53 &  Bw & AB  &      \\
 175 & 0103.1-7151 & 01 03 08.5        & -71 51 50                          & 16          & 2.67$\pm$1.36                              & 0.82$\pm$0.54  & 0.22$\pm$0.33     & &18&9& & & & && & A    &      \\
\noalign{\smallskip}
 176 & 0103.2-7242 & 01 03 15.7        & -72 42 37                          & 16          & 4.48$\pm$0.82                              & 0.74$\pm$0.22  & 0.45$\pm$0.15     & &59&15& & & & &&  Bw &   &      \\
 177 & 0103.2-7209 & 01 03 16.8        & -72 09 26                          & 18          & 9.05$\pm$0.88                              & 1.0            & 0.35$\pm$0.10     & 30 &107&24& 50 & 32 & & &  & R &    &rad snr\\
 178 & 0103.4-7219 & 01 03 24.8        & -72 19 16                          & 39          & 1.41$\pm$0.80                              & -0.20$\pm$0.53 & -1.0              & &10&18& & & & 16.1 & 88 &  F &    &      \\
 179 & 0103.4-7247 & 01 03 29.2        & -72 47 23                          & 12          & 9.59$\pm$1.07                              & 0.97$\pm$0.11  & 0.02$\pm$0.10     & &204&18& & & & 16.6 & 47 & A &     &      \\
 180 & 0103.7-7230 & 01 03 44.7        & -72 30 34                          & 15          & 1.50$\pm$0.50                              & 0.48$\pm$0.42  & 0.66$\pm$0.24     & &21&10& & & B0 & 14.8 & 87 &   &   &      \\
\noalign{\smallskip}
 181 & 0103.8-7254 & 01 03 53.3        & -72 54 49                          & 15          & 21.8$\pm$1.5                               & -1.0           &                   & &227&23& 72 & 13 & & &  &  Sl &  &      \\
 182 & 0103.9-7202 & 01 03 54.7        & -72 02 01                          & 11          & 1352$\pm$16                                & 0.93$\pm$0.01  & -0.11$\pm$0.01    & 25 &11073&36& 51 & 23 & B1I & 12.5 & 58 & R &    &rad snr\\
 183 & 0104.0-7201 & 01 04 04.2        & -72 01 56                          & 10          & 1885$\pm$13                                & 0.93$\pm$0.01  & -0.10$\pm$0.01    & 9 &19778&18& 51 & 21 & B1I & 12.5 & 79 &   &   &      \\
 184 & 0104.1-7244 & 01 04 08.8        & -72 44 04                          & 23          & 2.69$\pm$0.52                              & 1.0            & 0.57$\pm$0.18     & &25&13& & & & &&  Bw & AB  &      \\
 185 & 0104.2-7102 & 01 04 17.9        & -71 02 33                          & 95          &                                            &                &                   & 194 &32&43& & & & &&   &   &      \\
\noalign{\smallskip}
 186 & 0104.7-7203 & 01 04 46.6        & -72 03 57                          & 32          & 5.32$\pm$1.69                              & 1.0            & -0.42$\pm$0.53    & &31&22& & & F5I & 11.0 & 89 & A &     &      \\
 187 & 0104.7-7248 & 01 04 47.7        & -72 48 45                          & 18          & 1.33$\pm$0.57                              & 0.78$\pm$0.59  & 1.0               & &21&16& & & B0 & 14.0 & 79 &   &   &      \\
 188 & 0104.7-7253 & 01 04 47.8        & -72 53 53                          & 30          & 1.36$\pm$0.41                              & 1.0            & 0.71$\pm$0.25     & &13&21& & & & && Bw/D &  &      \\
 189 & 0105.0-7223 & 01 05 03.4        & -72 23 14                          & 11          & 240$\pm$4                                  & 0.66$\pm$0.01  & -0.32$\pm$0.01    & 44 &6430&32& 52 & 52 & B1 & 13.4 & 83 &  R &    &rad \\
 190 & 0105.3-7126 & 01 05 20.1        & -71 26 08                          & 88          & 54.5$\pm$5.4                               & -0.53$\pm$0.08 & 0.10$\pm$0.20     & &14&30& & & & &&  F &    &      \\
\noalign{\smallskip}
 191 & 0105.3-7210 & 01 05 20.4        & -72 10 21                          & 21          & 18.8$\pm$1.6                               & 1.0            & 0.07$\pm$0.09     & 57 &99&23& 53 & 58 & & &  &  R &    &rad \\
 192 & 0105.4-7219 & 01 05 28.4        & -72 19 36                          & 20          & 0.89$\pm$0.30                              & 0.99$\pm$0.82  & 0.07$\pm$0.34     & &11&14& & & & && & A     &      \\
 193 & 0105.7-7226 & 01 05 45.1        & -72 26 09                          & 24          & 1.22$\pm$0.35                              & 1.0            & 0.12$\pm$0.29     & &12&7& & & & && & A     &      \\
 194 & 0106.2-7243 & 01 06 12.3        & -72 43 38                          & 17          & 0.83$\pm$0.26                              & 1.0            & 0.34$\pm$0.32     & &14&10& & & BW & 14.6 & 56 & & A    &      \\
 195 & 0106.2-7205 & 01 06 15.1        & -72 05 25                          & 15          & 24.6$\pm$1.5                               & 1.0            & 0.24$\pm$0.06     & 42 &348&28& 54 & 49 & & & 20 &  R &    &      \\
\noalign{\smallskip}
 196 & 0106.3-7240 & 01 06 18.0        & -72 40 46                          & 18          & 0.47$\pm$0.21                              & 1.0            & 1.0               & &12&7& & & & &&  Bw &   &      \\
 197 & 0106.3-7125 & 01 06 19.8        & -71 25 40                          & 59          &                                             &                &                   & &27&34& & & & &&  D &  &      \\
 198 & 0106.6-7241 & 01 06 36.0        & -72 41 30                          & 18          & 0.76$\pm$0.25                              & 1.0            & -0.35$\pm$0.31    & &13&9& & & & && A &    &      \\
 199 & 0106.7-7220 & 01 06 43.9        & -72 20 45                          & 18          & 0.78$\pm$0.27                              & 1.0            & 0.68$\pm$0.38     & &12&13& & & & &&  Bw & A  &      \\
 200 & 0106.9-7224 & 01 06 57.7        & -72 24 55                          & 11          & 9.72$\pm$0.87                              & 0.87$\pm$0.08  & 0.45$\pm$0.08     & &426&10& 55 & 51 & & &  &  Bw &   &      \\
            \noalign{\smallskip}
            \hline
            \noalign{\smallskip}
\end{longtable}
\noindent
\clearpage
\vfill \eject

  \begin{longtable}[l]{rccccccccrrrcrrcclll}
   \hline 
  \noalign{\smallskip}
(1) &(2)          &(3)                 &(4)                                 &(5)          &(6)                       &(7)             &(8)                &(9)         &(10)   &(11)         &(12)  &(13)                   &(14)  &(15)   &(16)         &(17)   &(18) &(19) \\
    & Source      & RA                 & Dec                                &             & \multicolumn{1}{c}{Count Rate}              & \multicolumn{2}{c}{Hardness Ratio} &             &       &             & \multicolumn{2}{c}{Einstein} & \multicolumn{3}{c}{Simbad} &       &     \\
 No.& Name        & (J2000)            & (J2000)                            & $\rm P_{\rm  e}$ & Total                   & HR1            & HR2               & Ext         & LH    & $\Delta$    & ID   &  Dist                 & Type & Mag   & Dist        & Class & Class & Note\\
    & RX~J        & \p0h~\p0m~\p0s\p0~ & \p0\D\p0\p0$\arcmin$~\p0$\arcsec$  & ($\arcsec$) & \multicolumn{1}{c}{$\rm (10^{-3}\ s^{-1})$} &                &                   & ($\arcsec$) &       & ($\arcmin$) &      & ($\arcsec$)           &      &       &($\arcsec$)  & &(refined)      &     \\
            \noalign{\smallskip}
            \hline
            \noalign{\smallskip}
 201 & 0107.1-7231 & 01 07 09.1        & -72 31 03                          & 12          & 2.98$\pm$0.49            & 1.0            & 0.27$\pm$0.17     & &82&6& & & & 16.0 & 62 & & A    &      \\
 202 & 0107.1-7235 & 01 07 09.9        & -72 35 35                          & 15          & 0.73$\pm$0.26            & 1.0            & 1.0               & &18&6& 56 & 16 & & &  &  Bw &  &      \\
 203 & 0107.3-7248 & 01 07 18.5        & -72 48 32                          & 23          & 0.80$\pm$0.31            & 1.0            & 1.0               & &12&16& & & & &&  Bw &   &      \\
 204 & 0107.4-7243 & 01 07 27.4        & -72 43 21                          & 11          & 6.22$\pm$0.82            & 0.85$\pm$0.14  & 0.39$\pm$0.11     & &187&12& 82 & 31 & & &  &   &   &      \\
 205 & 0107.7-7226 & 01 07 44.7        & -72 26 01                          & 15          & 1.17$\pm$0.31            & 1.0            & 0.32$\pm$0.26     & &22&11& & & B0 & 13.8 & 41 & & A    &      \\
\noalign{\smallskip}
 206 & 0108.4-7228 & 01 08 27.2        & -72 28 56                          & 17          & 1.36$\pm$0.52            & 0.95$\pm$0.59  & 0.13$\pm$0.26     & &22&12& 59 & 61 & & &  & & A     &      \\
 207 & 0108.5-7247 & 01 08 31.1        & -72 47 12                          & 35          & 0.27$\pm$0.32            & 1.0            & 0.11$\pm$1.22     & &12&18& & & & && & AB     &      \\
 208 & 0108.5-7243 & 01 08 33.0        & -72 43 13                          & 21          & 1.21$\pm$0.37            & 1.0            & 0.39$\pm$0.30     & &14&15& & & & && & A    &      \\
 209 & 0108.6-7225 & 01 08 37.7        & -72 25 17                          & 15          & 5.41$\pm$0.84            & 0.84$\pm$0.19  & 0.47$\pm$0.12     & &89&15& 58 & 14 & & &  &  Bw & A &      \\
 210 & 0109.0-7243 & 01 09 02.9        & -72 43 59                          & 30          & 0.50$\pm$0.38            & 1.0            & 1.0               & 36 &13&17& & & & &&  Bw &   &      \\
\noalign{\smallskip}
 211 & 0109.0-7229 & 01 09 04.1        & -72 29 22                          & 17          & 2.94$\pm$0.69            & 0.80$\pm$0.31  & 0.33$\pm$0.17     & &42&15& & & & && & A    &      \\
 212 & 0109.4-7239 & 01 09 29.9        & -72 39 42                          & 22          & 1.18$\pm$0.34            & 1.0            & 0.24$\pm$0.28     & &15&17& & & & && & A    &      \\
 213 & 0109.5-7226 & 01 09 30.1        & -72 26 47                          & 27          & 1.87$\pm$0.59            & -0.30$\pm$0.27 & -0.06$\pm$0.44    & &10&18& & & & &&  F &   &      \\
 214 & 0110.5-7251 & 01 10 33.9        & -72 51 03                          & 20          & 0.53$\pm$0.20            & 1.0            & 0.93$\pm$0.38     & &12&13& & & & &&  Bw &   &      \\
 215 & 0110.6-7217 & 01 10 41.8        & -72 17 53                          & 40          & 2.10$\pm$0.87            & 0.80$\pm$0.60  & -0.16$\pm$0.28    & &12&27& & & & &&    &  &      \\
\noalign{\smallskip}
 216 & 0111.5-7302 & 01 11 35.5        & -73 02 01                          & 15          & 3.03$\pm$0.64            & 0.69$\pm$0.28  & 0.32$\pm$0.16     & &53&15& & & & && &  A  &rad \\
 217 & 0111.6-7321 & 01 11 39.5        & -73 21 27                          & 47          & 1.79$\pm$0.62            & 1.0            & 1.0               & &14&33& & & & &&  Bw &   &      \\
 218 & 0111.7-7250 & 01 11 43.1        & -72 50 25                          & 11          & 5.59$\pm$0.63            & 0.87$\pm$0.12  & 0.53$\pm$0.08     & &278&8& & & & &&  Bw & A  & quasar      \\
 219 & 0111.9-7255 & 01 11 56.1        & -72 55 20                          & 17          & 0.75$\pm$0.43            & 0.57$\pm$0.82  & 0.51$\pm$0.39     & &13&9& & & & && & AB     &      \\
 220 & 0111.9-7245 & 01 11 58.0        & -72 45 55                          & 15          & 1.08$\pm$0.43            & 0.47$\pm$0.51  & 0.54$\pm$0.27     & &21&7& & & & && & A    &      \\
\noalign{\smallskip}
 221 & 0111.9-7231 & 01 11 58.3        & -72 31 19                          & 23          & 1.52$\pm$0.36            & 1.0            & 0.51$\pm$0.25     & &20&19& 83 & 17 & & &  & & AB     &      \\
 222 & 0112.5-7301 & 01 12 30.3        & -73 01 40                          & 15          & 3.10$\pm$0.33            & 0.11$\pm$0.20  & -0.34$\pm$0.18    & &43&13& & & & &&  F &    &      \\
 223 & 0112.7-7207 & 01 12 43.6        & -72 07 25                          & 29          & 2.11$\pm$1.96            & 1.0            & 1.0               & 205 &265&41& & & & &&  R &    &      \\
 224 & 0112.8-7236 & 01 12 51.2        & -72 36 12                          & 14          & 2.58$\pm$0.39            & 0.92$\pm$0.31  & 0.25$\pm$0.16     & &61&13& & & & && & A     & quasar      \\
 225 & 0113.0-7244 & 01 13 03.4        & -72 44 41                          & 15          & 1.38$\pm$0.42            & 0.40$\pm$0.55  & 0.15$\pm$0.30     & &17&5& & & & &&      &      \\
\noalign{\smallskip}
 226 & 0113.1-7241 & 01 13 06.2        & -72 41 36                          & 13          & 1.40$\pm$0.28            & 1.0            & -0.03$\pm$0.20    & &44&8& & & & && A &     &      \\
 227 & 0113.7-7239 & 01 13 43.6        & -72 39 50                          & 15          & 0.74$\pm$0.21            & 1.0            & 0.90$\pm$0.20     & &25&9& & & & &&  Bw &   &      \\
 228 & 0114.1-7250 & 01 14 08.2        & -72 50 47                          & 14          & 1.43$\pm$0.45            & 0.09$\pm$0.33  & 0.48$\pm$0.26     & &23&4& & & & &&  F &    &      \\
 229 & 0114.1-7258 & 01 14 08.4        & -72 58 15                          & 16          & 0.63$\pm$0.20            & 1.0            & 0.25$\pm$0.33     & &14&10& & & & && & A    &      \\
 230 & 0114.2-7234 & 01 14 12.5        & -72 34 51                          & 23          & 1.00$\pm$0.28            & 1.0            & 0.34$\pm$0.29     & &13&15& & & & && & AB     &      \\
\noalign{\smallskip}
 231 & 0114.2-7319 & 01 14 12.5        & -73 19 43                          & 34          & 2.96$\pm$0.64            & 1.0            & 0.71$\pm$0.23     & &27&31& 85 & 29 & & &  &  Bw & H  &rad \\
 232 & 0114.7-7303 & 01 14 43.3        & -73 03 02                          & 22          & 1.49$\pm$0.35            & 1.0            & 0.61$\pm$0.24     & &24&15& & & & &&  Bw & AB  &      \\
 233 & 0114.8-7249 & 01 14 51.9        & -72 49 44                          & 18          & 0.0$\pm$0.41             & 0.83$\pm$1.12  & 0.67$\pm$0.31     & &14&7& & & & &&  & AB    &      \\
 234 & 0115.0-7244 & 01 15 00.1        & -72 44 52                          & 13          & 0.14$\pm$0.28            & 1.0            & 0.71$\pm$0.17     & &55&8& & & & &&  Bw &   &      \\
 235 & 0115.1-7254 & 01 15 06.6        & -72 54 08                          & 15          & 0.73$\pm$0.21            & 1.0            & -0.25$\pm$0.28    & &17&9& & & & && A &     &      \\
\noalign{\smallskip}
 236 & 0115.1-7257 & 01 15 09.6        & -72 57 55                          & 14          & 1.98$\pm$0.50            & 0.38$\pm$0.31  & 0.57$\pm$0.20     & &39&12& & & & &&   &   &      \\
 237 & 0115.9-7237 & 01 15 54.3        & -72 37 46                          & 20          & 0.83$\pm$0.26            & 1.0            & 0.32$\pm$0.31     & &11&16& & & & && & AB     &      \\
 238 & 0116.5-7259 & 01 16 35.9        & -72 59 41                          & 17          & 4.98$\pm$0.59            & 1.0            & 0.43$\pm$0.11     & 24 &86&18& 62 & 27 & & &  &  Bw & A  & quasar      \\
 239 & 0116.6-7240 & 01 16 38.0        & -72 40 18                          & 15          & 0.39$\pm$0.53            & 0.80$\pm$0.36  & 0.64$\pm$0.17     & &48&17& & & & 14.5 & 18 &    &  &      \\
 240 & 0116.7-7246 & 01 16 42.6        & -72 46 24                          & 15          & 1.66$\pm$0.51            & 0.57$\pm$0.40  & 0.61$\pm$0.23     & &31&15& & & & && & A    &      \\
\noalign{\smallskip}
 241 & 0116.8-7314 & 01 16 52.0        & -73 14 55                          & 48          & 2.27$\pm$0.58            & 1.0            & 0.16$\pm$0.28     & &12&30& & & & &&  D &  &      \\
 242 & 0117.0-7326 & 01 17 03.3        & -73 26 28                          & 11          & 271$\pm$3                & 1.0            & 0.31$\pm$0.01     & 32 &8494&41& 63 & 20 & & &  &  Bl &   &      \\
 243 & 0117.2-7238 & 01 17 14.7        & -72 38 02                          & 30          & 0.51$\pm$0.47            & -1.0           &                   & &12&20& & & & && Sw/D & &      \\
 244 & 0119.1-7256 & 01 19 06.7        & -72 56 49                          & 24          & 3.96$\pm$0.62            & 1.0            & 0.74$\pm$0.17     & &51&26& & & & &&  Bw &   &      \\
 245 & 0119.4-7301 & 01 19 26.7        & -73 01 19                          & 19          & 25.4$\pm$1.4             & 1.0            & 0.62$\pm$0.06     & 62 &324&29& 66 & 11 & & &  &  R &    &      \\
\noalign{\smallskip}
 246 & 0120.2-7245 & 01 20 17.0        & -72 45 20                          & 37          & 5.81$\pm$1.16            & -0.15$\pm$0.19 & 0.06$\pm$0.26     & &14&31& & & A9/F0 & 7.8 & 26 &  F &    &star  \\
 247 & 0121.4-7258 & 01 21 24.8        & -72 58 28                          & 31          & 7.49$\pm$0.96            & 1.0            & 0.28$\pm$0.14     & &54&37& & & B0 & 14.4 & 89 & & A     &      \\
 248 & 0121.4-7244 & 01 21 28.2        & -72 44 13                          & 37          & 5.14$\pm$0.88            & 1.0            & 0.71$\pm$0.20     & &33&36& & & & & &  Bw &   &      \\
            \noalign{\smallskip}
            \hline
            \noalign{\smallskip}
\end{longtable}
}
\noindent
\clearpage
\vfill \eject

\end{document}